\documentclass[acmtog]{acmart}
\acmSubmissionID{276}

\usepackage{booktabs} 

\usepackage[linesnumbered, ruled]{algorithm2e} 

\SetAlFnt{\small}
\SetAlCapFnt{\small}
\SetAlCapNameFnt{\small}
\SetAlCapHSkip{0pt}

\acmConference[SIGGRAPH '25]{Make sure to enter the correct
  conference title from your rights confirmation email}{June 03--05,
  2018}{Woodstock, NY}

\acmISBN{978-1-4503-XXXX-X/18/06}

\acmJournal{TOG}

\usepackage{amsmath,amsfonts,bm}

\def\eqref#1{equation~\ref{#1}}

\def\1{\bm{1}}

\def\rva{{\mathbf{a}}}

\def\rvc{{\mathbf{c}}}
\def\rvd{{\mathbf{d}}}

\def\rvh{{\mathbf{h}}}

\def\rvp{{\mathbf{p}}}
\def\rvq{{\mathbf{q}}}

\def\rvs{{\mathbf{s}}}

\def\rvx{{\mathbf{x}}}

\DeclareMathAlphabet{\mathsfit}{\encodingdefault}{\sfdefault}{m}{sl}
\SetMathAlphabet{\mathsfit}{bold}{\encodingdefault}{\sfdefault}{bx}{n}

\def\gC{{\mathcal{C}}}
\def\gD{{\mathcal{D}}}

\def\gL{{\mathcal{L}}}
\def\gM{{\mathcal{M}}}

\newcommand{\textroot}{\text{root}}
\newcommand{\textjoints}{\text{joints}}

\newcommand{\ddpmE}{\mathbb{E}_{\rvx_0, \gC \sim D} \mathbb{E}_{k \sim p(k)} \mathbb{E}_{\rvx_k \sim q(\rvx_k|\rvx_0)}}

\copyrightyear{2025}
\acmYear{2025}
\setcopyright{acmlicensed}\acmConference[SIGGRAPH Conference Papers '25]{Special Interest Group on Computer Graphics and Interactive Techniques Conference Conference Papers }{August 10--14, 2025}{Vancouver, BC, Canada}
\acmBooktitle{Special Interest Group on Computer Graphics and Interactive Techniques Conference Conference Papers (SIGGRAPH Conference Papers '25), August 10--14, 2025, Vancouver, BC, Canada}
\acmDOI{10.1145/3721238.3730616}
\acmISBN{979-8-4007-1540-2/2025/08}

\begin{document}

\title{PARC: Physics-based Augmentation with Reinforcement Learning for Character Controllers}

\author{Michael Xu}
\email{mxa23@sfu.ca}
\affiliation{%
  \institution{Simon Fraser University}
  \city{Burnaby}
  \state{British Columbia}
  \country{Canada}
}

\author{Yi Shi}
\email{ysa273@sfu.ca}
\affiliation{%
  \institution{Simon Fraser University}
  \city{Burnaby}
  \state{British Columbia}
  \country{Canada}
}

\author{KangKang Yin}
\affiliation{%
  \institution{Simon Fraser University}
  \city{Burnaby}
  \state{British Columbia}
  \country{Canada}
}

\author{Xue Bin Peng}
\affiliation{%
  \institution{Simon Fraser University, NVIDIA}
  \city{Burnaby}
  \state{British Columbia}
  \country{Canada}
}

\begin{abstract}
Humans excel in navigating diverse, complex environments with agile motor skills, exemplified by parkour practitioners performing dynamic maneuvers, such as climbing up walls and jumping across gaps. Reproducing these agile movements with simulated characters remains challenging, in part due to the scarcity of motion capture data for agile terrain traversal behaviors and the high cost of acquiring such data.
In this work, we introduce PARC (\textbf{P}hysics-based \textbf{A}ugmentation with \textbf{R}einforcement Learning for Character \textbf{C}ontrollers), a framework that leverages machine learning and physics-based simulation to iteratively augment motion datasets and expand the capabilities of terrain traversal controllers. PARC begins by training a motion generator on a small dataset consisting of core terrain traversal skills. The motion generator is then used to produce synthetic data for traversing new terrains. However, these generated motions often exhibit artifacts, such as incorrect contacts or discontinuities. To correct these artifacts, we train a physics-based tracking controller to imitate the motions in simulation. The corrected motions are then added to the dataset, which is used to continue training the motion generator in the next iteration.
PARC’s iterative process jointly expands the capabilities of the motion generator and tracker, creating agile and versatile models for interacting with complex environments. 
PARC provides an effective approach to develop controllers for agile terrain traversal, which bridges the gap between the scarcity of motion data and the need for versatile character controllers.
\end{abstract}



\begin{teaserfigure}
  \includegraphics[width=\textwidth]{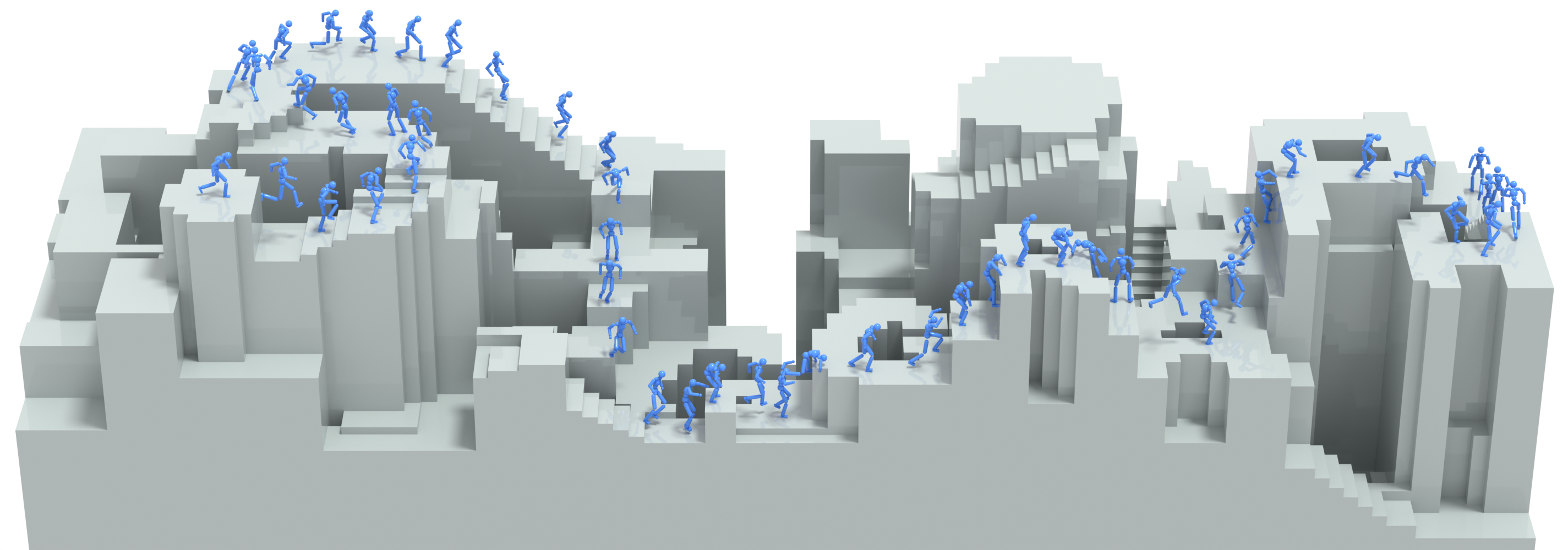}
  \caption{A visualization of a physically simulated character traversing a complex terrain using different agile parkour skills. Our framework starts with a small dataset of terrain traversal motions, and then performs an iterative dataset augmentation loop that progressively expands the capabilities of a motion generation model and a physics-based motion tracking controller.}
  \Description{Heightmap parkour agent.}
  \label{fig:teaser}
\end{teaserfigure}


\maketitle

\section{Introduction}
Humans possess the remarkable ability to navigate through diverse and complex environments by employing a broad range of agile motor skills. A prime example of human agility can be seen in parkour practitioners, who regularly demonstrate extraordinary athleticism 
by stylishly traversing obstacles using dynamic 
combinations of maneuvers, such as vaulting, climbing, and jumping. Developing simulated characters that can achieve comparable versatility remains a significant challenge. Current state-of-the-art methods 
for training physics-based controllers predominantly rely on motion capture data, using imitation objectives to guide the learning process towards natural and human-like behaviors.
Due to the challenges of capturing highly athletic interactions with complex environments, there is little data available of human athletes traversing diverse terrains with agile motor skills.
Moreover, acquiring large quantities of high quality motion data for such athletic behaviors can be exorbitantly expensive.

Human motion capture data is important for training character controllers that can move in a natural and life-like manner. 
While recording a large volume of data from human athletes can be costly, it is feasibly economical to capture a narrow set of high quality motion data depicting core terrain-traversal skills.
A simple data augmentation method leveraging generative models may be to first train a motion generator on this small initial dataset, then use the generative model to automatically generate an expanded synthetic dataset of more diverse behaviors. This augmented dataset can be used to further train the motion generator to improve its generation capabilities for traversing more difficult and diverse environments.
This \emph{self-consuming} process can be performed iteratively to create a sufficiently large dataset suitable for training a general and effective terrain traversal controller. 

However, a motion generator trained on a small dataset often produces low-quality, physically implausible motions, especially in unfamiliar scenarios. These artifacts—such as incorrect contacts, floating, sliding, or discontinuities—can degrade the generator’s performance over the iterations. To mitigate this progressive degradation, we leverage physics simulation by training a motion imitation controller to correct these artifacts. Instead of using the raw generated motions, the physically corrected motions are added to the dataset, improving the physical realism of the synthetic data and stabilizing the training process throughout the iterative process. 

Building on this idea, we introduce PARC (Physics-based Augmentation with Reinforcement Learning for Character Controllers), a framework that takes a small initial motion dataset as input and outputs a motion tracking controller for traversing complex terrains. 
PARC iteratively trains a motion generator and motion tracker to augment a motion dataset, progressively expanding the capabilities of both models. The motion generator synthesizes new terrain traversal motions, while the motion tracker refines them to ensure physical plausibility before adding the motion clips to the dataset. This expanded dataset is used to continuously train both the generator and tracker, enhancing their versatility. Through multiple iterations, PARC develops an expressive motion generator and an agile motion-tracking controller. Together, these components enable precise control of a simulated character, allowing it to agilely navigate complex obstacle-filled environments. The code and data used to train PARC, as well as the models and data generated by PARC, can be found at \url{https://github.com/mshoe/PARC}.

\section{Related Work}

Recent advances in machine learning have led to a surge of techniques that are capable of automatically producing high-fidelity human motions, which span both kinematics-based methods \citep{holden2017phase, starke2019neural, rempe2021humor} and physics-based methods \citep{peng2018deepmimic, peng2021amp, peng2022ase, 2021-supertrack, yao2022cvae}. 
However, generating highly dynamic character-scene interaction behaviors remains a persistent challenge due to limited data availability. The constraints imposed by complex scenes also place more stringent demands on motion quality, as they are susceptible to more pronounced artifacts, such as floating and terrain collisions. In this section, we review the most relevant prior work on generating interactive motion under imposed terrain constraints.

\subsection{Kinematic Motion Generation}
Given abundant, high-quality motion data, kinematic motion generation models can effectively synthesize complex human behaviors.
\citet{holden2017phase, holden2020learned} introduced a phase-based auto-regressive model that is able to generate locomotion behaviors on irregular terrain. 
\citet{yi2024humanscene, li2023controi} leveraged the expressiveness of diffusion models to synthesize scene-aware motions. While these works have shown promising results, they often struggle to generalize to new scenarios not captured in the original training dataset. Furthermore, their lack of physics-based simulation often leads to artifacts such as floating, ground penetration, and self-collision, compromising the realism of the synthesized motions.

\subsection{Physics-based Character Control}
Physics-based character simulation has been explored as a means to procedurally generate novel behaviors in scenarios where motion data may be scarce. For example, to model human athletic skills, existing works have developed physics-based controllers capable of replicating a wide range of dynamic sports, including parkour \citep{2012-terrain-runner}, tennis \citep{2023-tennis}, table tennis \citep{wang2024strategy}, soccer \citep{xie2021soccer}, boxing \citep{won2021box}, basketball \citep{wang2024skillmimic, liu2018basketball}, and climbing \citep{ClilmbingNaderi2017}. 
However, many of the methods introduced in these studies rely heavily on imitating existing motion data, thereby limiting their applicability in domains where high-quality motion data is scarce or all together unavailable.

Methods that combine physics-based and kinematic methods can leverage the advantages of these two paradigms to further enhance motion quality, diversity, and generalization, while also mitigating artifacts that violate physical principles. \citet{kevin2019drecon} employed motion matching as a kinematic planner and trained a physics-based controller to track the planner's motions. \citet{jiang2023drop} utilized generative motion priors and projective dynamics for natural character behaviors. \citet{yuan2023physdiff} applied physics-based tracking to refine motion diffusion outputs. These approaches rely on high-quality motion datasets for training their kinematic and tracking models. In contrast, our method iteratively trains a motion generator and tracker using an initially small dataset, expanded with physics-corrected motions.

\subsection{Motion Control for Terrain Traversal}
Developing motor controllers capable of agile traversal across complex terrains has been an active area of research spanning multiple fields, from robotics to computer graphics.
\citet{2012-terrain-runner} trained controllers specialized for traversing different types of obstacles and then employed manually designed planners to sequence these skills, enabling the traversal of sequences of different obstacles.
\citet{2017-deep-loco} proposed a hierarchical reinforcement learning framework for training controllers capable of complex locomotion tasks such as ball dribbling across terrains, trail following, and obstacle avoidance. 
\citet{yu2021human} presented an algorithm that produces a control policy and scene arrangement to imitate dynamic terrain-traversal motions from video.

In the field of robotics, there is a large body of work that applied reinforcement learning methods to train controllers that enable legged robots to traverse through environments with obstacles using agile locomotion skills \citep{hoeller2024anymal, zhang2024learningagilelocomotionrisky}. While some techniques can generate highly dynamic locomotion skills without relying on demonstrations or reference motion data, the resulting controllers are prone to producing unnatural behaviors.

\subsection{Data Augmentation}
Data augmentation has been an essential technique to prevent overfitting and improve generalization since the advent of the deep learning \cite{krizhevsky2012imagenet, shorten2019augsurvey}. Data augmentation has also been a vital tool in many character animation frameworks \cite{park2019learning, holden2017phase}. 
More recently, researchers have explored self-consuming generative models as a more powerful methodology to automate data augmentation.
\citet{gillman2024selfcorrecting} showed that self-consuming motion diffusion models experience mode collapse, unless a self-correcting function is used. To mitigate model collapse when training on self generated data, \citet{gillman2024selfcorrecting} incorporated a pre-trained physically simulated motion tracking controller \cite{luo2021dynamics} as a correction function for physically implausible motion artifacts. 
While the previous work uses physics-based motion trackers as correction functions, our work integrates the motion tracker as a component of the self-consuming loop by continually training the tracker to correct the synthetic motions produced by the motion generator. Our work also applies the self-consuming, self-correcting loop to the challenging task of terrain traversal.

\section{Background}
In this section, we review the core machine learning concepts underlying our framework. First, we discuss diffusion models, the primary architecture used for the motion generators. Next, we cover reinforcement learning, the paradigm used to train physically simulated motion tracking controllers.

\subsection{Diffusion Models}
A generative model is trained to generate samples from an unknown data distribution, which is approximated with a dataset of samples $\gD$.
Diffusion models are a type of generative model which have recently been shown to be effective for motion synthesis \cite{tevet2022human}.
A diffusion model learns to generate samples from a data distribution $\gD$ by learning to reverse a diffusion process. 
A diffusion process takes a sample $\rvx_0$ from $\gD$, and slowly converts it to a sample from a standard Gaussian distribution by iteratively applying noise to the sample. 
At the final diffusion timestep $K$, the distribution converges to a standard Gaussian $x_K \sim \mathcal{N} (0, \mathbf{I})$.  
While the original DDPM formulation trains a denoising model to predict the noise applied to the original data sample \cite{ho2020denoising}, many motion diffusion models choose to train a denoising model $G$ that directly predicts the denoised motion sample instead \citep{tevet2022human, cohan2024flexible}. Please refer to \citet{karunratanakul2023gmd} for a more detailed discussion between the two options.
The denoising model is trained using a simple reconstruction objective:
\begin{equation}
\label{eq:simpleL}
    \gL_{\text{rec}}(G) := \ddpmE \left [ || \rvx_0 - G (\rvx_k, k, \mathcal{C})||^2 \right ],
\end{equation}
where $p(k)$ represents the diffusion timestep distribution (e.g. uniform distribution between $[1, K]$), and $\gC$ is a context associated with the sample $\rvx_0$ (e.g. text, control signal, etc.). 
After training, a reverse diffusion process can be applied to generate samples using the denoising model.

\subsection{Reinforcement Learning}
Reinforcement learning has been an effective paradigm for developing controllers for a wide range of tasks \cite{duan2016benchmarkingdeepreinforcementlearning, 2017-deep-loco}. 
In reinforcement learning, an agent interacts with its environment according to a policy $\pi$ to maximize an objective \citep{2018-sbrl}. At each time step $t$, the agent observes the state of the environment $s_t$. The agent then samples and executes an action from a policy $\rva_t \sim \pi (\rva_t | \rvs_t)$. The next state $\rvs_{t+1}$ is then determined by the environment dynamics $\rvs_{t+1} \sim p(\rvs_{t+1} | \rvs_t, \rva_t)$. After every state transition, the agent receives a scalar reward determined by a reward function $r_t = r(\rvs_t, \rva_t, \rvs_{t+1})$. The agent's objective is to learn a policy that maximizes its expected discounted return $J(\pi)$,

\begin{equation}
    J(\pi) = \mathbb{E}_{p(\tau | \pi)} \Biggl [ \sum_{t=0}^{T-1} \gamma^t r_t \Biggr],
\end{equation}
where $p(\tau | \pi)$ represents the likelihood of a trajectory $\tau$ under policy $\pi$, $T$ denotes the time horizon of a trajectory, and $\gamma \in [0, 1]$ is a discount factor.

\section{System Overview}
\begin{figure}
    \centering
    \centerline{\includegraphics[width=1.1\columnwidth]{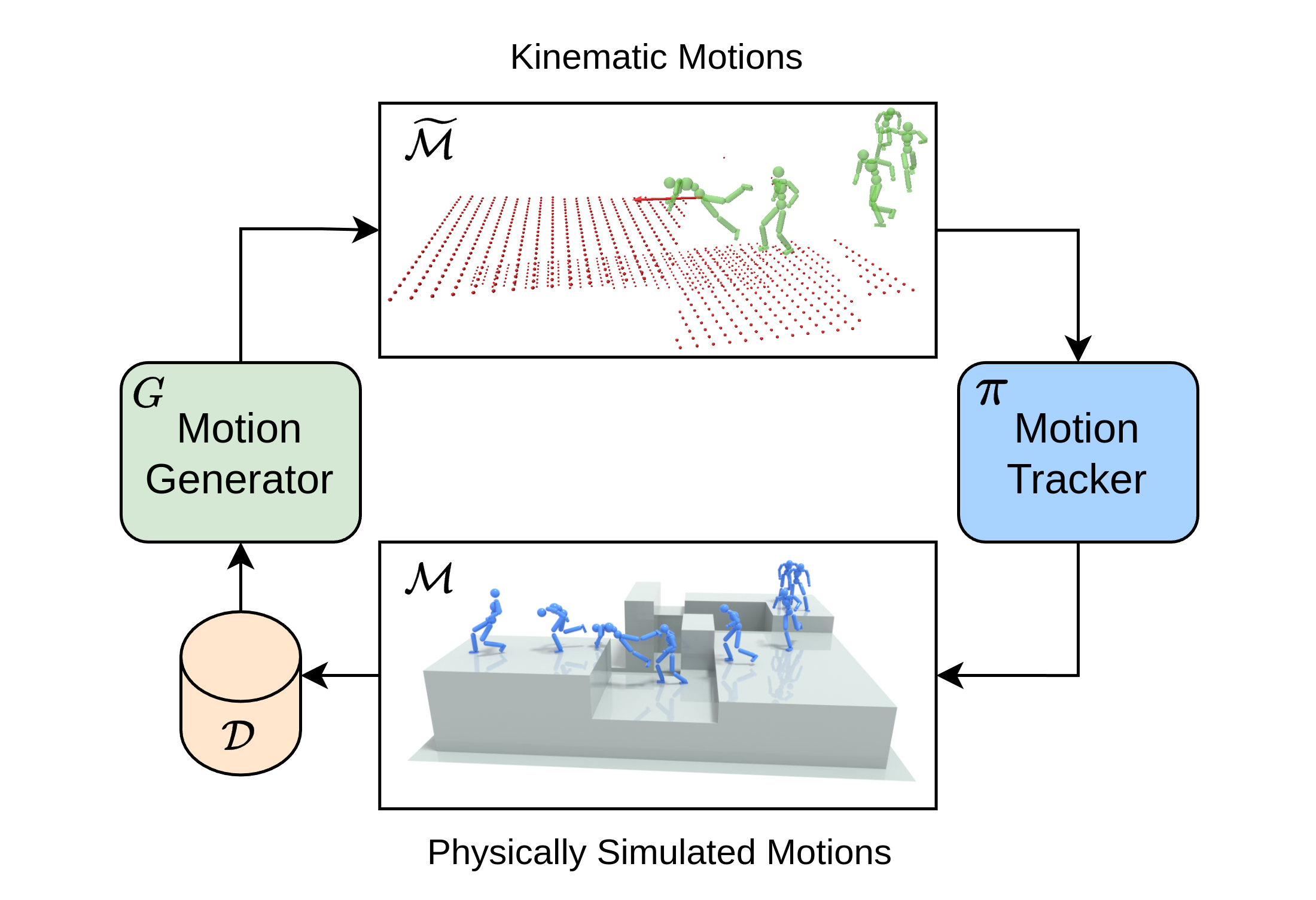}}
    \caption{Overview of the PARC framework. PARC iteratively trains a motion generator and motion tracker with self-generated motion data. The motion generator produces kinematic motion sequences to train the motion tracker, while the motion tracker corrects physics-related artifacts in a simulator, enabling the motion generator to continue training on new physics-based motions.}
    \label{fig:method-high}
\end{figure}

In this work, we present PARC (\textbf{P}hysics-based \textbf{A}ugmentation with \textbf{R}einforcement Learning for Character \textbf{C}ontrollers), an iterative data augmentation framework for training agile terrain traversal controllers for physically simulated characters. An overview of PARC is available in Figure \ref{fig:method-high}. 
Our framework consists of two main components, a motion generator that generates kinematic motions given a target terrain, and a physics-based motion tracker that corrects artifacts in the generated motions by leveraging physical simulation.
These two components are applied iteratively to augment the motion dataset, progressively expanding the versatility of the generator and capabilities of the tracker.

PARC starts with a small initial motion-terrain dataset $\gD^0$.
At each iteration $i$, 
PARC trains a motion generation model $G^i$ 
on the dataset $\gD^{i-1}$ from the previous iteration. Next, $G^i$ is used to synthesize new motions $\widetilde{\gM}^i$ for traversing new terrains. 
After synthesizing new motions, a motion tracking policy $\pi^i$ is trained to enable a simulated character to imitate the motions from the combined dataset $\gD^{i-1} \cup \widetilde{\gM}^i$. 
Once trained, $\pi^i$ is used to record physically corrected motions 
$\gM^i$ by tracking $\widetilde{\gM}^i$ in simulation. Finally, the physically corrected motions are then added to the dataset $\gD^i \gets \gD^{i-1} \cup \gM^i$.

An important characteristic 
of PARC is that the models are trained in a continual manner. 
The generator $G^i$ and policy $\pi^i$ of each stage is initialized with the trained models from the previous stage, utilizing past experience to accelerate the learning of new motions.
The final motion generation model can be used to synthesize target motions on new terrains, and the physics-based motion tracker can then follow the target motion to control a physically simulated character to traverse new environments. 

\section{Motion Generator}
The motion generator is one of the main components of PARC's self-augmentation loop, and acts as a planner that generates kinematic motions for traversing a given terrain.
The motion generator is represented as a diffusion model \cite{ho2020denoising} trained to generate motion sequences while conditioned on a local terrain heightmap and target direction. An overview of our network architecture is shown in Figure \ref{fig:mdm-arch}.
Given 
an input context $\mathcal{C}$, 
the motion generator predicts a motion sequence $\rvx = \{ \rvx^1, \rvx^2, \cdots, \rvx^{N} \}$ for traversing the terrain along the desired direction.
Each frame $\rvx^i$ of the motion sequence $\rvx$ is represented using a set of features consisting of:
\begin{itemize}
    \item $\rvp^0 \in \mathbb{R}^3$, root position 
    \item $\rvq^0 \in \mathbb{R}^3$, root rotation 
    \item $\rvq \in \mathbb{R}^{J \times 3}$, joint rotations
    \item $\rvp \in \mathbb{R}^{J \times 3}$, joint positions
    \item $\rvc \in [0, 1]^{J}$, contact labels
\end{itemize}
where $J$ denotes the number of joints in the character's body.
All rotations are represented with exponential maps.


The input context $\mathcal{C}$ to the diffusion model consists of a
heightmap $\rvh$, recorded in the character's local coordinate frame,
the horizontal target direction $\rvd \in \mathbb{R}^2$,
and the first two frames of the motion sequence. These first two frames are an optional condition, allowing our model to both generate an initial motion sequence given no previous frames and also generate long-horizon motion sequences autoregressively. Conditioning on two input frames, instead of one, provides the model with velocity information. 

The diffusion model is implemented  with a transformer encoder architecture, similar to MDM \citep{tevet2022human}.
An illustration of the model architecture is available in Figure \ref{fig:mdm-arch}.
The generator $G$ receives as input the context, $\mathcal{C} = \{ \rvh, \rvd, \rvx^{1}, \rvx^{2}\}$, the noisy motion frames $\rvx_k$, and diffusion timestep $k$. The generator then predicts the clean motion sequence $\hat{\rvx}_0$,
\begin{equation}
    G(k, \rvx_k, \mathcal{C}) = \hat{\rvx}_{0} = \{\hat{\rvx}_0^1, \hat{\rvx}_0^2, \cdots, \hat{\rvx}_0^{N}\}
\end{equation}
The input diffusion timestep $k$, heightmap $\rvh$, target direction $\rvd$, and noisy frames $\rvx_k$ are first encoded using different embedding networks to map each into tokens for the transformer.
The heightmap observations are processed with a convolutional neural network, and the image patches are extracted and processed into tokens by an MLP. 
The target direction is encoded with an MLP to produce one token, and each frame of the input motion sequence $\rvx_k$ is encoded into a token with an MLP. 
When the first two frames $(\rvx_0^1, \rvx_0^2)$ are given, they replace the frames $(\rvx_k^1, \rvx_k^2)$ in the output sequence, and are encoded with the same encoding network.
Positional encoding is applied to all tokens.
The output token sequence is passed through a final MLP that maps it to the denoised motion $\hat{\rvx}_0$.

\subsection{Motion Data Sampling}
\begin{figure}
    \centering
    \includegraphics[width=1.0\linewidth]{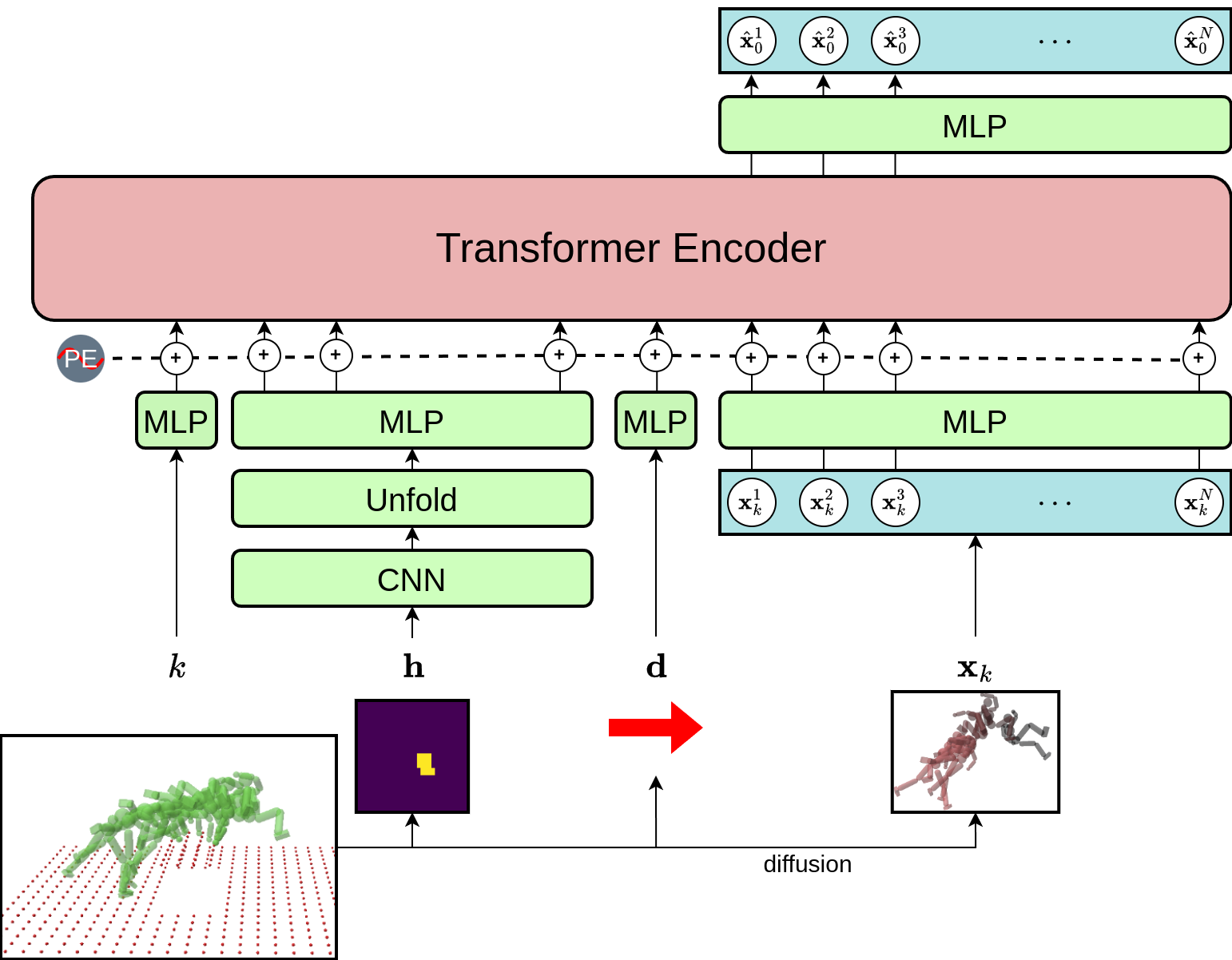}
    \caption{
    The transformer encoder based architecture of the terrain-conditioned motion generator. $\rvh$ is first processed by a CNN into an image of shape 64x16x16, then unfolded into 64 non-overlapping image patches of shape 64x2x2. The image patches are then embedded into tokens with an MLP. The target direction $\rvd$ is embedded into a single token with an MLP. Each frame of the noisy motion sequence $\rvx_k$ is embedded into a token using an MLP.}
    \label{fig:mdm-arch}
\end{figure}

The motion generator is trained to generate motions for traversing new terrains using a dataset of motion clips paired with their respective terrains. When a motion clip is sampled, a half-second motion sequence is selected uniformly from the frames of the motion clip. 
Figure \ref{fig:mdm-arch} illustrates the features that are extracted from each sequence. 
 The motion sequence is split into 13 future frames and 2 past frames. Each frame within a sequence is canonicalized relative to the second frame, which is treated as the most recent frame that the generator is conditioning on. 
A random future frame is used to determine the target direction. 


The sampling of the local heightmap from the global terrain geometry associated with the motion clip is done using a \( 31 \times 31 \) uniform grid of points, canonicalized to the second frame of the motion. The sampled heightmaps are augmented with randomly oriented boxes with varying heights to improve generalization when given out of distribution heightmap observations. 
A collision-avoidance technique is employed to ensure these boxes do not add terrain penetration to the motion, as detailed in Appendix \ref{app:terrain}.

\subsection{Training}
The motion generator is trained following the standard DDPM process illustrated  in Equation \ref{eq:simpleL} , but with additional geometric loss terms, which are introduced to better capture the spatial coherence and physical plausibility of generated motions.
Training is done by iteratively sampling $\rvx_k$ for $k \sim [1, K]$, predicting $\hat{\rvx}_0$, and computing a diffusion model loss. The training loss is given by:
\begin{equation}
\label{eq:mdm-loss}
    \gL(G) = \gL_{\text{rec}}(G) + \gL_{\text{velocity}}(G) + \gL_{\text{joint}}(G) + \gL_{\text{pen}}(G),
\end{equation}
where $\gL_{\text{rec}}$ is the reconstruction loss, $\gL_{\text{velocity}}$ is a velocity loss,
$\gL_{\text{joint}}$ is a joint consistency loss, and $\gL_{\text{pen}}$ is a terrain penetration loss. 
Details of these losses are provided in Appendix \ref{app:mdm-losses}.

\subsection{Terrain-Aware Motion Generation}
\label{sec:terrain-aware-mgen}
One of the core challenges for training a terrain-conditioned motion generator is ensuring the generated motions respect the surrounding terrain, either by not penetrating the terrain, or by employing physically plausible motor skills to interact with the terrain. When trained only on the small initial dataset $\gD^0$, $G$ tends to produce motions that ignore the physical constraints imposed by the terrain such as running into a wall instead of climbing over it. 
We hypothesize that when generating a motion autoregressively, $G$ is overfitting to the previous frames as a consequence of using a small dataset. 
This focus on the previous frame results in the motion generator ignoring the terrain condition, leading to motions that fail the comply with the surrounding terrain.
Our first technique to address this issue is to train $G$ with an auxiliary motion-terrain penetration loss $\mathcal{L}_{\text{pen}}$, details of which are available in Appendix \ref{app:losses}. 
To further improve the generated motion's compliance with a given terrain, during training with 10\% probability we use the generator to synthesize motions on random terrains and apply the terrain penetration loss to resolve intersections with the terrain  while also disabling the reconstruction loss. This approach encourages the model to avoid terrain penetration, even when confronted with unexpected variations in the terrain.

Incorporating terrain-penetration loss during training encourages the motion generator to better adhere to the surrounding terrain. However, we found that results could still be improved by using an additional technique to further enhance terrain compliance. From our hypothesis that motion-terrain penetration is a result of overfitting to previous frames, we blend the output of our model when conditioned with and without the previous frames. This approach is similar to classifier-free guidance \cite{2022-cfg}, and provides a tradeoff between temporally smooth motions with respect to the previous frames, and terrain-compliant motions. Motions with smoothness artifacts can be corrected by the physics-based motion tracking controller, while motions that severely violate terrain constraints, such as running through a wall, cannot be reproduced in simulation. The blended denoising update is determined by:
\begin{equation}
\label{eq:cfg}
    \begin{split}
        G_{\text{blend}}\left(k, \rvx_k, \gC=\left(\rvh, \rvd, \rvx_0^1, \rvx_0^2\right)\right) \\ 
        = s G\left(k, \rvx_k, \gC =(\rvh, \rvd)\right) + (1-s)G\left(k, \rvx_k, \gC=\left(\rvh, \rvd, \rvx_0^1, \rvx_0^2\right)\right),
    \end{split}
\end{equation}
where \( s \) is the blending coefficient. We found that \( s=0.65 \) works well. To faciliate this blending during inference, \( G(k, \rvx_k, \gC = \{\rvh, \rvd \}) \) is trained simultaneously with \( G(k, \rvx_k,  \gC = \left\{\rvh, \rvd, \rvx_0^1, \rvx_0^2\right\}) \) by randomly masking the attention to the two previous frame tokens \( \rvx_0^1 \) and \( \rvx_0^2 \) with a 15\% chance. At test time, evaluating the unconditional and conditional models is done by supplying the corresponding attention mask to the inputs.

\subsection{Synthesizing New Motions}
Given a terrain and a path, our trained model can generate long motion sequences by conditioning autoregressively on its own output frames. We use this in combination with procedurally generated terrains to synthesize new motions at each iteration. While a motion tracking controller can correct physics-based artifacts in a reference motion, low quality motions may be too challenging for the tracker to follow. Therefore, two techniques are incorporated to reduce artifacts in the generated motions.
First, motions are generated in batches of 64 sequences at a time, and a heuristic loss is used to select the best motion within the batch. The heuristic loss combines a contact loss, penetration loss, and path incompletion penalty. Once a motion has been selected within a batch, the motion is refined using kinematic optimization techniques to mitigate artifacts such as jittering, terrain penetration, and contact consistency. More details on the kinematic correction procedure are available in Appendix \ref{app:kmc}.

\begin{figure*}
    \centering
    \includegraphics[width=0.49\linewidth]{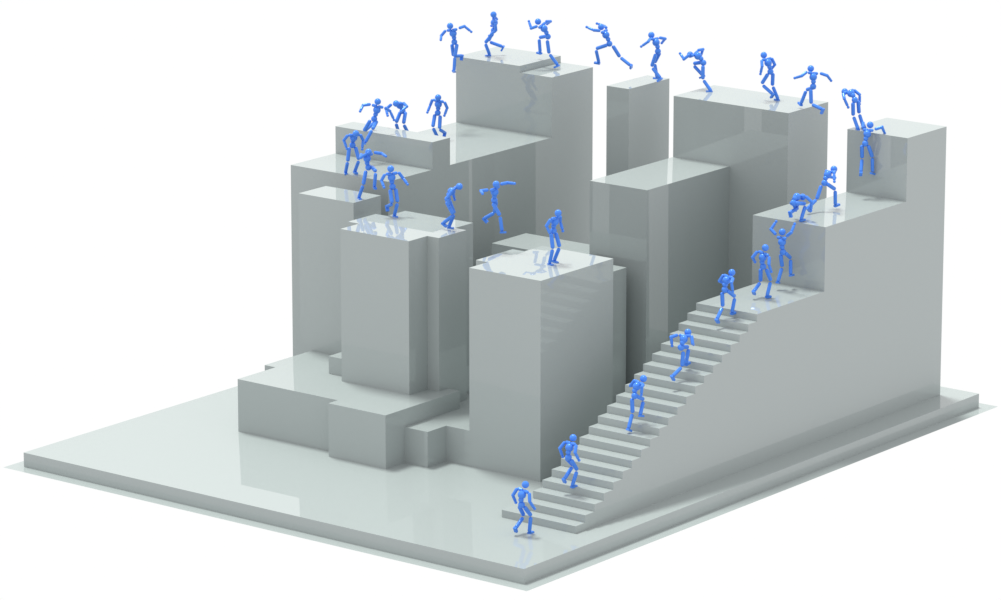}
    \includegraphics[width=0.49\linewidth]{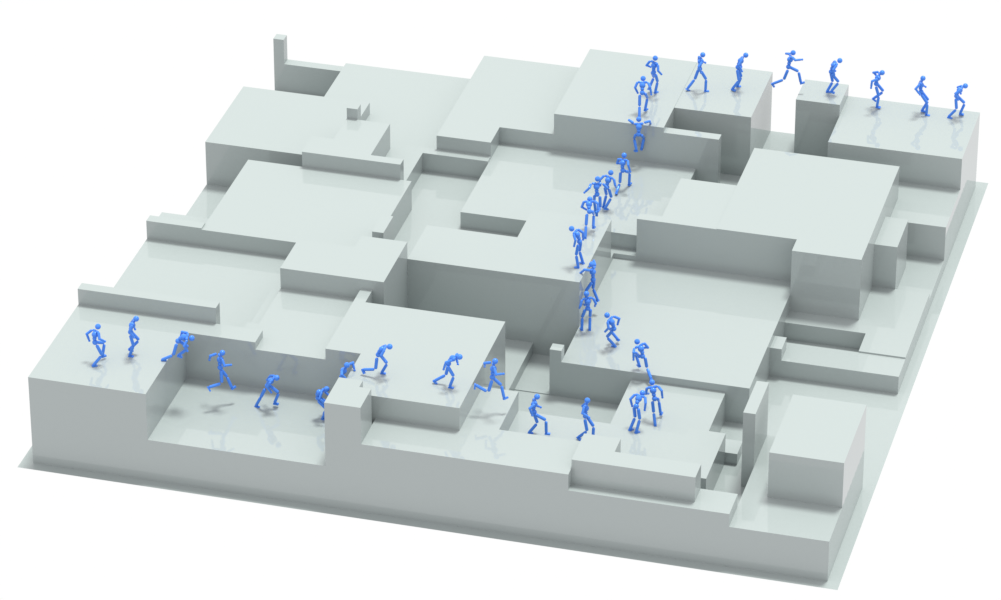}
    \includegraphics[width=0.49\linewidth]{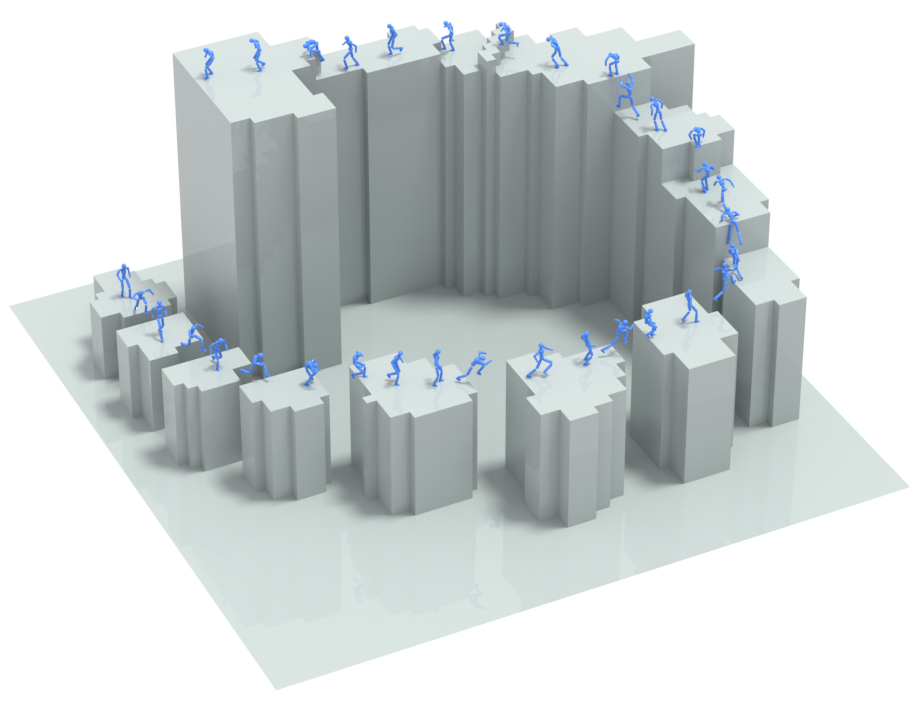}
    \includegraphics[width=0.49\linewidth]{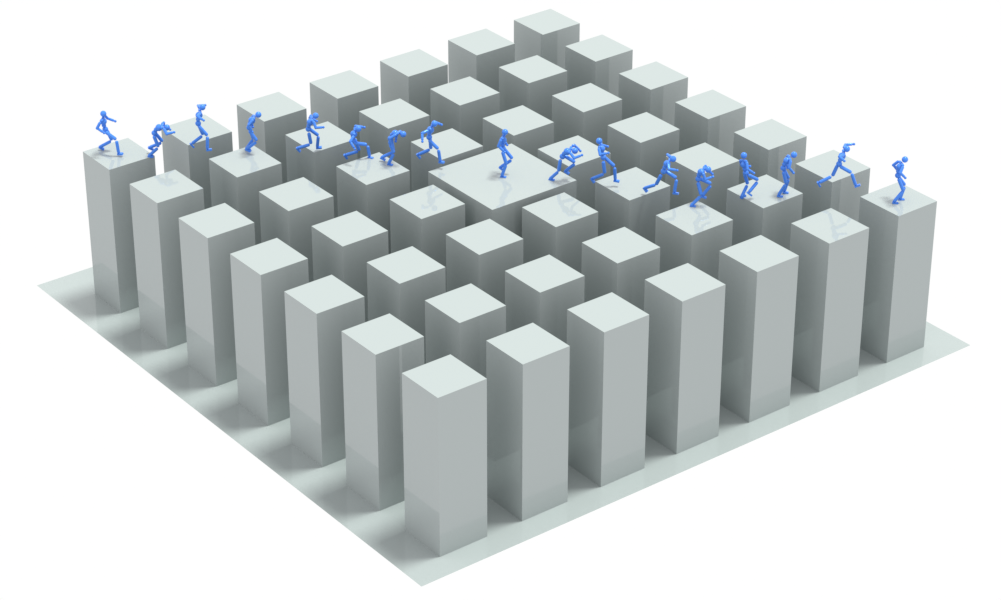}
    \caption{Long-horizon physics based motions generated using the final motion generator and motion tracker of PARC. 
    }
    \label{fig:results}
\end{figure*}

\section{Tracking Controller}
In this section, we detail the process for training a controller capable of executing the generated kinematic motion sequences in simulation to enable a physically simulated character to traverse complex environments. The character is controlled using a motion tracking controller, trained with reinforcement learning on the kinematically generated motions.
Our motion tracking controller follows the DeepMimic framework \cite{peng2018deepmimic}, and all physics simulations are performed using Isaac Gym \cite{makoviychuk2021isaac}.

\subsection{Observation and Action Representation}
The actions from the policy specifies the target orientations for PD controllers positioned at each joint. The policy is queried at 30$Hz$, while the physics simulation is performed at 120$Hz$.
The observations of the agent consist of its proprioceptive state, its local terrain observations, and future target frames from the reference motion. The proprioceptive state consists of the agent's root position $\rvp^{\textroot}$, root rotation $\rvq^{\textroot}$, joint rotations $\rvq^{1:N_{\textjoints}}$, joint positions $\rvp^{1:N_{\textjoints}}$  and contact labels $\rvc^{1:N_{\textjoints}}$. The local terrain observations are represented using a heightmap of points $\rvh$ sampled around the character. The target states $\rvs^{\text{ref}}$ are specified by the reference motion clip, and all features are canonicalized with respect to the character's local frame. The local frame of the character is defined with the origin at the root, the x-axis facing the root link's facing direction, and the z-axis aligned with the global up vector.

\subsection{Training}
The physics-based controller is trained with a DeepMimic-inspired motion-tracking objective, enabling it to navigate diverse terrains by replicating the kinematic motions generated by the motion generator. The tracking reward is designed to encourage the controller to minimize the differences between the agent and reference motion's root position, root velocity, joint rotations, joint velocities, key body positions, and contact labels. The contact label is a binary signal that specifies whether a body is in contact with the environment. We found that matching the reference motion's contacts is vital for ensuring the simulated character interacts with an environment using naturalistic contact configurations. A detailed description of the reward function is available in Appendix \ref{rl-rewards}. The policy network is represented with three fully connected layers with 2048, 1024, and 512 units, and is trained using Proximal Policy Optimization \cite{2017-ppo},
with advantages computed using $\text{GAE}(\lambda)$ \cite{schulman2015high}. The value function is trained using target values computed with $\text{TD}(\lambda)$ \cite{2018-sbrl}.

\subsection{Physics-Based Motion Correction}
Once the tracking controller is trained in each iteration, it is utilized to generate physics-corrected versions of the kinematically produced motions. These corrected motions, recorded within the simulation, effectively reduce physics-related artifacts present in the original kinematic motions. The successfully recorded motions are subsequently added to the motion dataset, which is then used to continue training the motion generator in the next iteration of PARC.
A recorded motion is considered unsuccessful if the character fails to reach the end of the motion. Some reference motions may have large artifacts or challenging frames in their early frames, making it difficult to track. To help increase the output of successful recorded motions from our tracking controller, we initialize the character at different times along the motion, and then the earliest initialization that successfully reaches the final frame will be recorded as additional motion data.

\section{Experiments}
\begin{figure}[b]
    \centering
    \includegraphics[width=0.8\linewidth]{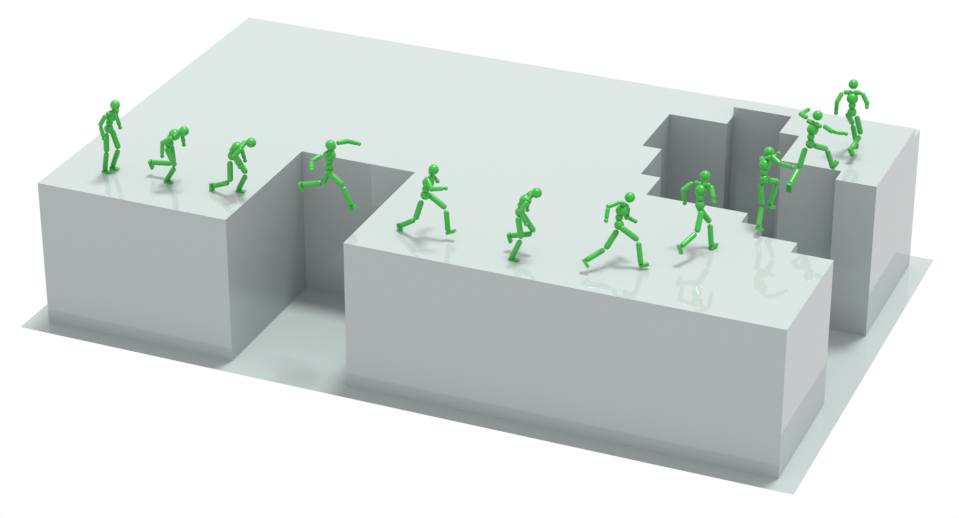}
    \includegraphics[width=0.8\linewidth]{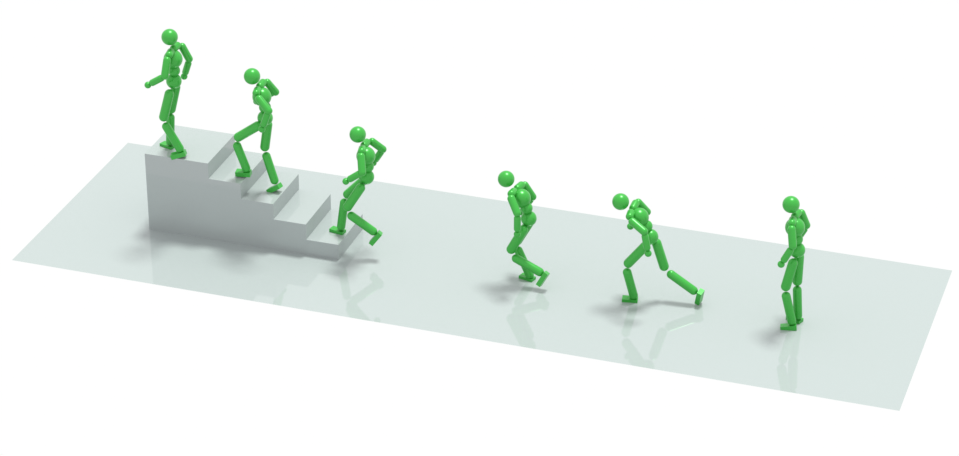}
    \includegraphics[width=0.8\linewidth]{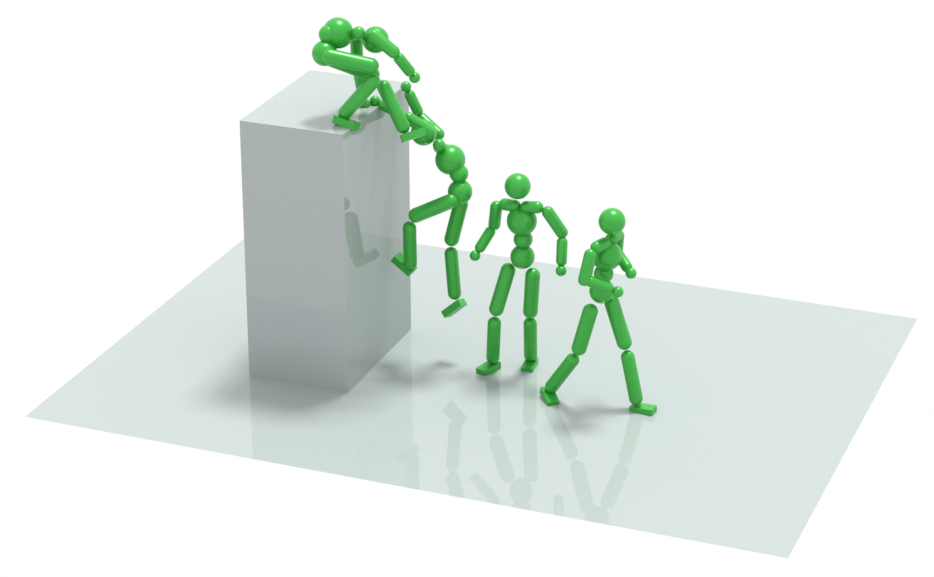}
    \caption{Examples of terrain-traversal motions found in our original dataset. The terrain is typically very simple, and the vast majority of clips focus on showcasing one particular parkour skill such as jumping (top), running up stairs (middle), and climbing walls (bottom).}
    \label{fig:og-dataset}
\end{figure}

Our initial parkour dataset consists of new  motion capture data for various terrain traversal skills such as vaulting, climbing, jumping, and running. 5.5 seconds of motion clips are also recorded from the Unreal Engine Game Animation Sample Project \cite{2024unreal}. The corresponding terrains for each motion clip are manually reconstructed to fit each motion. The contact labels in the original dataset are manually labeled.
The original dataset contains a total of 14 minutes and 7 seconds of motion data, depicting parkour skills such as climbing, vaulting, running on flat and bumpy ground, moving on and off platforms, and going up and down stairs. Examples of motions in the original dataset can be seen in Figure \ref{fig:og-dataset}.

Since the original motion dataset is relatively small, there is a severe lack of spatial diversity, with most terrain-traversal skills being performed on terrains with fixed heights. To improve the spatial variations of the dataset, we applied random adjustments to the terrains, and used manually-designed heuristics to adapt the motions to the adjusted terrains. To improve the physical plausibility of the adjusted motions, we also trained a physics-based motion tracking controller on this augmented dataset and then recorded the motions performed by the physically simulated character. The simulation environment is able to detect and label accurate contact labels automatically. This process is repeated to generate 50 spatial variations of each motion clip in the original dataset, increasing the coverage of terrain variations for our input dataset to the PARC framework.
 The initial synthetic data expansion does take advantage of the generative capabilities of the motion generator to discover new skills, and is only used to improve the motion generator in the initial iteration of PARC. 

\begin{figure}
\centering
    \includegraphics[width=0.49\linewidth]{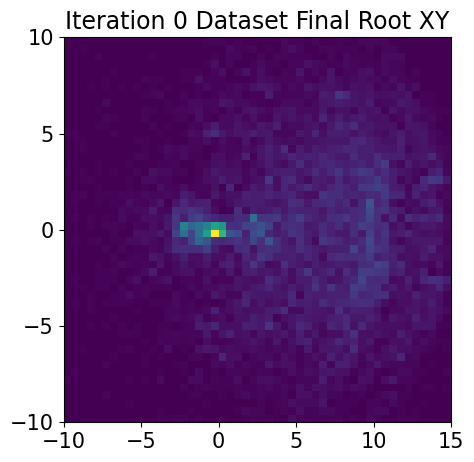}
    \includegraphics[width=0.49\linewidth]{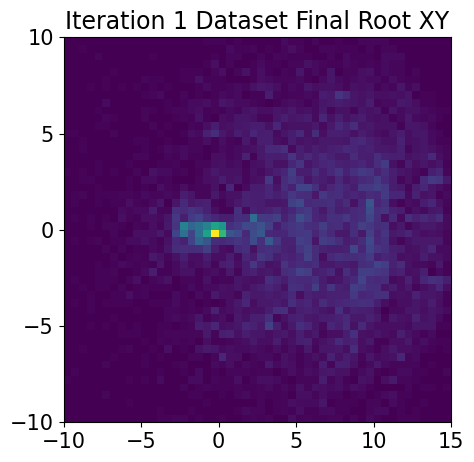}
    \includegraphics[width=0.49\linewidth]{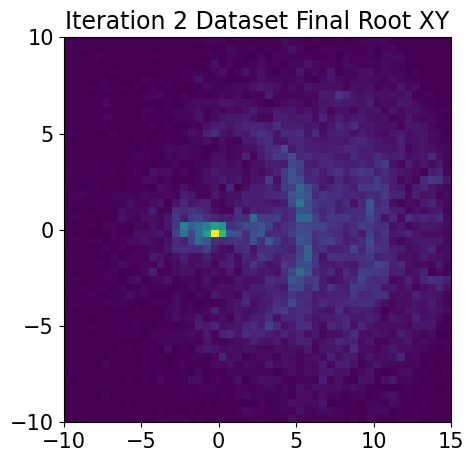}
    \includegraphics[width=0.49\linewidth]{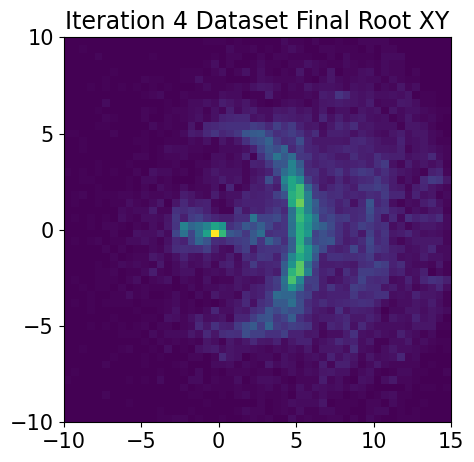}
    \caption{ A visualization of the distribution of the final relative horizontal (XY) root positions from motion clips in the dataset as at different PARC iterations. As the PARC iterations increase (left to right, top to bottom), the dataset expands and increases the diversity of trajectories.
    }
    \label{fig:kde-parc}
\end{figure}

Given the initial dataset, the PARC framework is applied for three iterations using a single A6000 GPU, requiring approximately one month to complete. 
In the first two iterations, the motion generator is used to generate approximately 1000 new motions on randomly generated terrains. 
In both the third and fourth iterations, the motion generator is used to generate 2000 motions on a large manually designed terrain, more details of which are available in Appendix \ref{app:terrain}. 
All generated motions have a maximum length of 10 seconds.
 A visualization of the dataset expansion can be seen in Figure \ref{fig:kde-parc}, where the net root horizontal displacement of every motion clip is plotted as a distribution. 
The initial dataset contained many forward-moving motions, but as training progressed, trajectories became more diverse, covering a wider range of directions.

\subsection{Novel Behaviors}
\begin{figure}[b]
    \centering
    \includegraphics[width=0.8\linewidth]{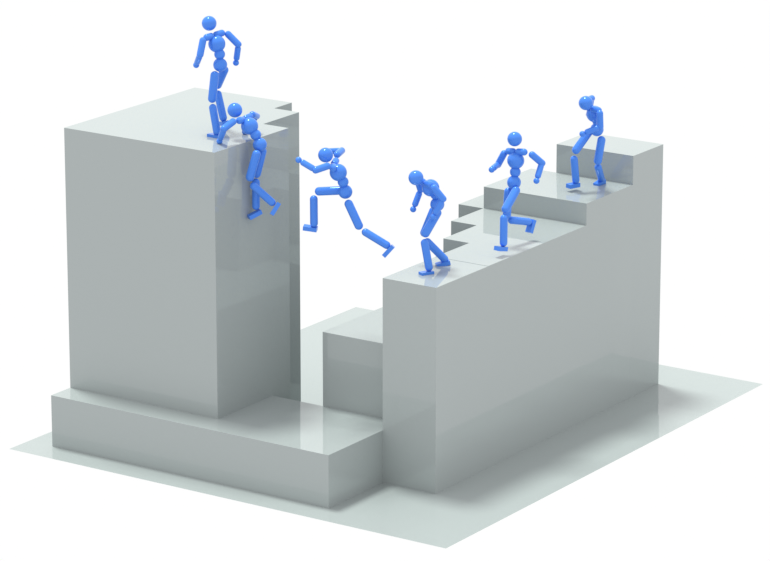}
    \includegraphics[width=0.8\linewidth]{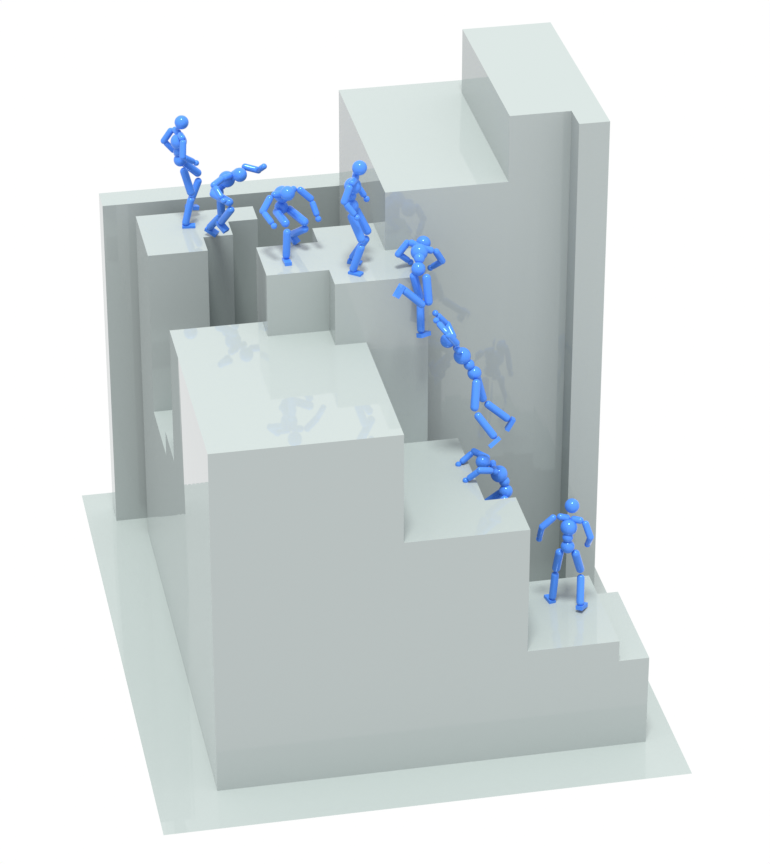}
    \includegraphics[width=0.8\linewidth]{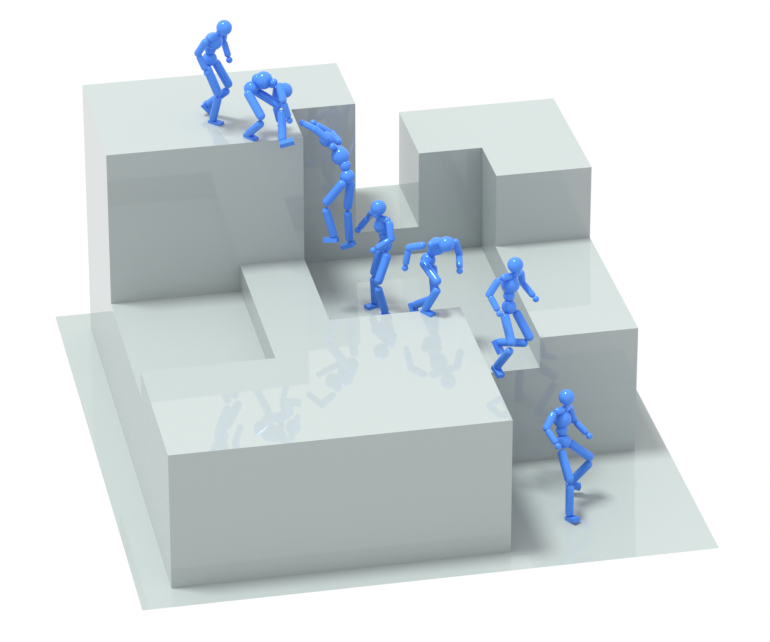}
    \caption{Examples of novel physics-based motions generated by PARC. (Left) A character combines a jumping motion with a climbing motion to catch onto a higher ledge. (Middle) A character first jumps a gap, then holds onto a ledge and drops. While falling, the character uses their hands to catch onto another ledge before landing. (Right) A character climbs down and then runs on and off a platform, landing on a lower ground level.}
    \label{fig:novel-motions}
\end{figure}

Our PARC framework can generate novel behaviors that go beyond those in the original dataset, enabling traversal of significantly more complex and diverse terrains.
Examples of new behaviors generated by our framework are shown in Figure \ref{fig:novel-motions}. The physically simulated character is able to sequence together different skills such as jumping across a gap and catching onto a ledge.
Furthermore, the models can be used to generate significantly longer motions for traversing complex terrains. First, given a target path and a large terrain, the motion generator is used to autoregressively generate a kinematic target motion. Then, the motion tracker follows the target motion to traverse across the terrain.
Examples of behaviors produced by the simulated character are shown in Figure \ref{fig:results}. 
The time required to generate each 0.5 seconds of motion with a batch size of 32 is approximately 12 seconds on an A6000 GPU.  
For the long-horizon examples, a batch size of 32 is used and heuristic criteria are applied to automatically select the best motion (see Appendix \ref{app:kmc}). 

\begin{table}[h]
\centering
\caption{Quantitative results of our motion generators across PARC iterations with the best values bolded. These metrics measure various aspects of motion quality, and include FWD (final waypoint distance), TPL (terrain penetration loss), TCL (terrain contact loss), and \%HJF (percentage of high jerk frames). The motion generators from each iteration are used to generate 32 motions for each of the 100 test terrains. The average value across all 3200 generated test motions is reported for each metric.}
\begin{tabular}{|c | c c c c|}
 \hline
 \textbf{Iteration} & \textbf{FWD} $\downarrow$ & \textbf{TPL} $\downarrow$ & \textbf{TCL} $\downarrow$ & \textbf{\%HJF} $\downarrow$ \\ [0.35ex] 
 \hline\hline
 1 & 1.908 & 2093 & 114.1 & 10.70 \\ 
 \hline
 2 & 1.586 & 705.5 & 9.761 & 4.387 \\
 \hline
 3 & 0.747 & 448.2 & \textbf{8.070} & 3.238 \\
 \hline
 4 & \textbf{0.596} & \textbf{179.6} & 9.763 & \textbf{2.730} \\
 \hline
 no physics correction & 1.572 & 547.3 & 17.44 & 18.68 \\
 \hline
\end{tabular}
\label{tbl:mdm-metrics}
\end{table}

\subsection{Motion Generator Performance}
\begin{figure}
    \centering
    \includegraphics[width=0.49\linewidth]{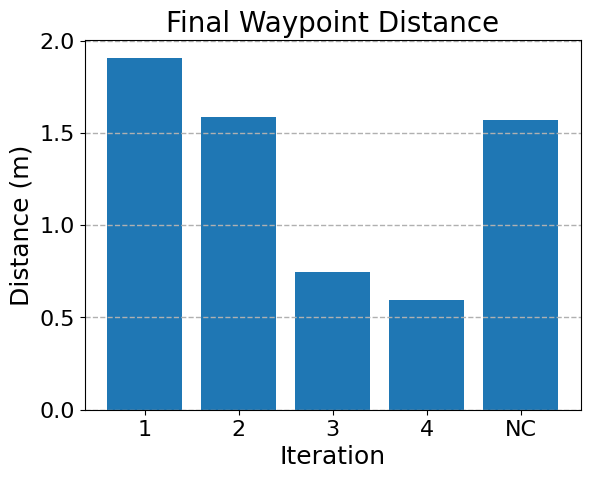}
    \includegraphics[width=0.49\linewidth]{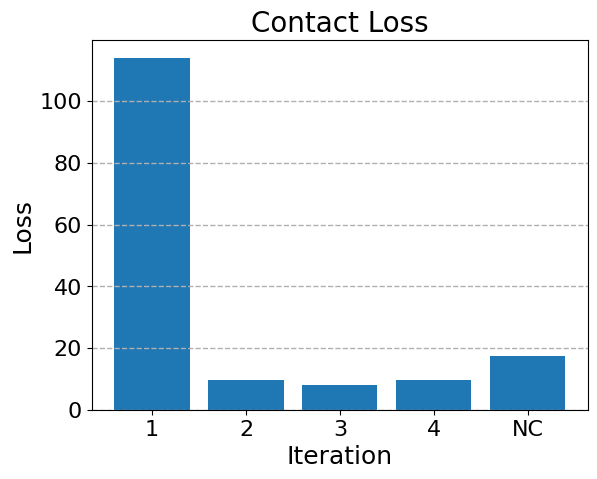}
    \includegraphics[width=0.49\linewidth]{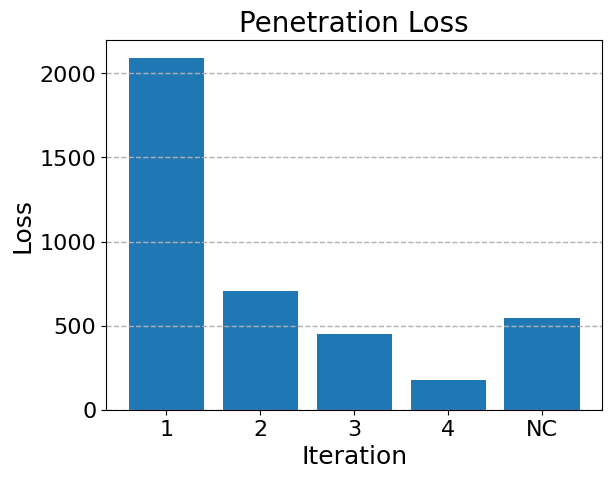}
    \includegraphics[width=0.49\linewidth]{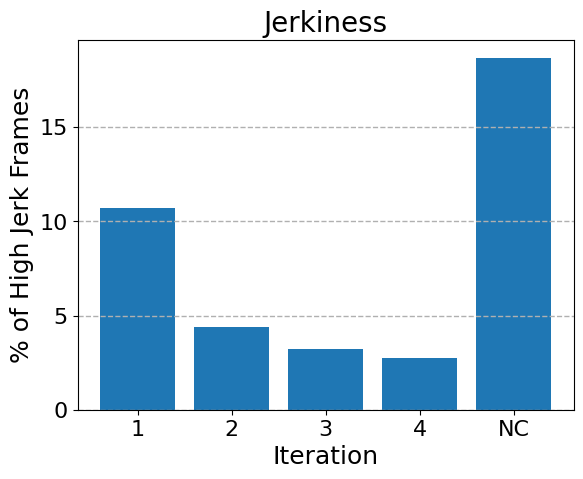}
    \caption{Plots showing the measured quantitative results of generated motions from the kinematic motion generator across different PARC iterations, including an iteration with no physics-based motion correction (labeled "NC"). Each metric reports the mean calculated over 3200 motions that were generated across 100 different test terrains for each motion generator. Without physics-based correction, the models generate motions that are much less physically realistic.}
    \label{fig:mdm-metrics}
\end{figure}

To evaluate the improvements to the motion generator from each PARC iteration, we conducted a quantitative experiment evaluating a large number of generated motions.
We created 100 new test terrains and target paths using a procedural terrain generation algorithm different from the one used for the augmented dataset.
Next, we applied various iterations of our PARC motion generator to produce 32 motions for each terrain, resulting in a total of 3200 motions for each PARC iteration.
Blended denoising is applied with a coefficient of $s=0.65$, and DDIM with a stride of 5 is used to generate motions along the target paths. No kinematic or physics-based motion corrections are applied to the generated kinematic motions. Four heuristic metrics are used to evaluate the quality of the generated motions: final waypoint distance (FWD), terrain penetration loss (TPL), terrain contact loss (TCL), and percentage of high jerk frames (\%HJF). 
The final waypoint distance measures the distance between the root position of the last generated frame and the target position at the end of a path. High jerk frames are determined as any frame where the jerk of any joint is greater than the maximum joint jerk observed in the original mocap dataset, which is approximately 11666 $m/s^3$. Table \ref{tbl:mdm-metrics} and Figure \ref{fig:mdm-metrics} summarize the performance of the different models. The metrics exhibit significant improvements as the number of PARC iterations progresses. 
To demonstrate the importance of physics-based correction, we include an experiment where we trained a motion generator using uncorrected motions from the first PARC iteration (labeled "no correction").
We use the selection heuristic (described in Appendix \ref{app:kmc}) to filter motions for terrain penetration and contact losses, with no other correction technique to highlight the importance of the physics-based motion tracking stage.

\begin{table}[h]
\centering
\caption{Quantitative results of our motion tracker for different PARC iterations. The success rate is the tracker's average rate of motion completion over 100 generated test motions. The joint tracking error is computed using an average of 2048 episodes for each of the 100 generated test motions at random initial timesteps.}
\begin{tabular}{|c | c c|}
 \hline
 \textbf{Iteration} & \textbf{Success Rate (\%) $\uparrow$} & \textbf{Joint Tracking Error (m) $\downarrow$} \\ [0.35ex] 
 \hline\hline
 1 & 27 & 0.08294 \\ 
 \hline
 2 & 44 & 0.05851 \\
 \hline
 3 & 60 & 0.05321 \\
 \hline
 
 4 & \textbf{68} & \textbf{0.05167} \\
 \hline
\end{tabular}
\label{tbl:dm-metrics}
\end{table}

\begin{figure}
    \centering
    \includegraphics[width=0.9\linewidth]{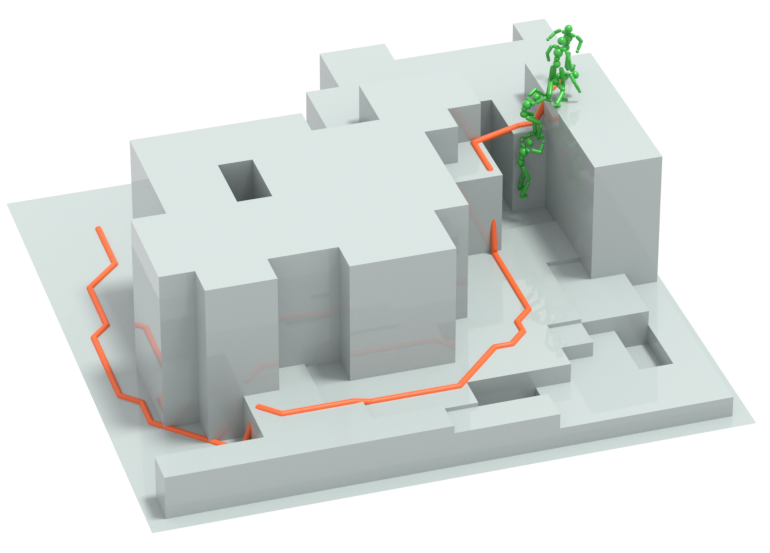}
    \includegraphics[width=0.9\linewidth]{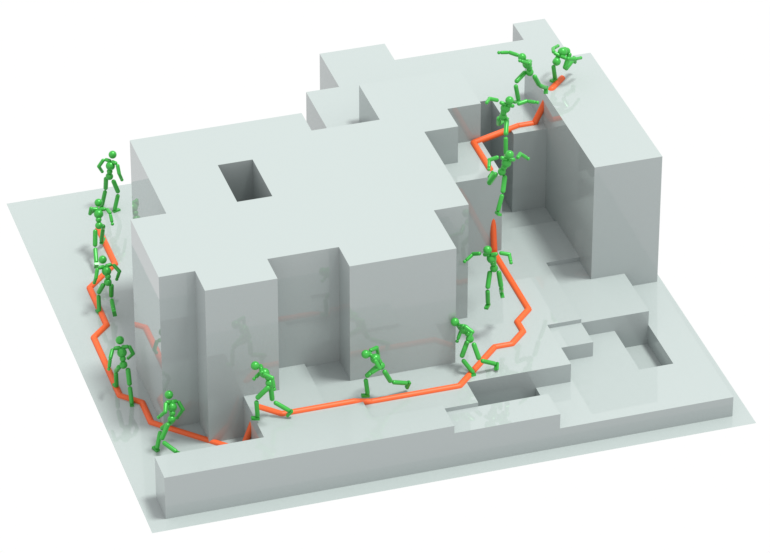}
    \includegraphics[width=0.9\linewidth]{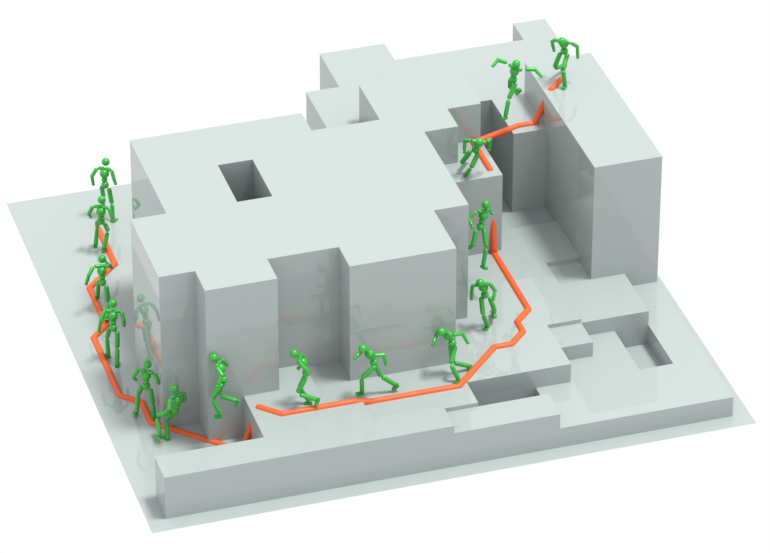}
    \caption{Motions generated on a test terrain for different iterations of PARC. Each motion was generated using a batch of 32 for up to 15 seconds of motion time and then automatically selected based on a heuristic incorporating terrain penetration, contact loss, and incompletion penalty. (Left) The iteration 1 motion generator is only trained on the initial dataset, and struggles to navigate across complex terrain. The character was only able to get off the cliff within 15 seconds. (Middle) The motion produced by a motion generator trained on uncorrected generated data from the iteration 1 motion generator. It exhibits physically implausible artifacts such as changing directions while flying through the air. (Right) The motion generated by the iteration 3 generator shows the character utilizing contacts with the terrain to navigate to the end of the path.}
    \label{fig:mdm-test-examples}
\end{figure}

The motion generator trained with uncorrected motions produces significantly more jerk related artifacts, which can be seen in the \%HJF metric. Qualitatively, these motions tended to include physically impossible feats, such as changing the trajectory of the character in the middle of a jump, which is likely due to the motion generator being trained on kinematically generated motions with no physics-based correction.
Figure \ref{fig:mdm-test-examples} compares the behavior of the motion generator from different iterations on a challenging example. The motion generator from the final iteration (Iteration 3) is able to generate higher quality and more physically realistic motions that successfully follow the target path across the environment.

\subsection{Motion Tracker Performance}
To evaluate the performance of the tracking controllers from different PARC iterations, we assess their success rates and average joint errors when tracking the test motions generated by their corresponding motion generator. The success rate is the percentage of episodes where the motion tracking controller is able to reach the final frame of motion when starting from the first frame. The average joint error is the average distance between the joints of the simulated character and the corresponding joints in the reference motion, and is computed using the average of 2048 episodes with random initial frames for each motion.
We report these metrics in Table \ref{tbl:dm-metrics}.
The generated motions come from the previous experiment on motion generator performance, except we use the selection heuristic (Appendix \ref{app:kmc}) to first select the best motion generated for each test terrain. Therefore, the motion tracking controllers are tested with 100 generated target motions.


\subsection{Blended Denoising}
To demonstrate the importance of the blended denoising as described in Section \ref{sec:terrain-aware-mgen}, we experimented with different values for $s$ in the blended denoising update in Equation \ref{eq:cfg}. To do this, we used the iteration 4 motion generator to generate 3200 test motions on our test terrain set. Lower values of the blending coefficient tend to produce smoother motions as indicated by the percentage of high jerk frames metric, but perform much worse in terms of terrain penetration and terrain contact losses, as seen in Table \ref{tbl:cfg-metrics}.

\begin{table}[h]
\centering
\caption{Quantitative results of our 4th iteration motion generator using different blending coefficients $s$. These metrics measure various aspects of motion quality, and include FWD (final waypoint distance), TPL (terrain penetration loss), TCL (terrain contact loss), and \%HJF (percentage of high jerk frames). The motions with the best quality have a balance between terrain compliance (low FWD, TPL, TCL) and temporal continuity (low \%HJF). We used $s=0.65$ for automatically augmenting the dataset through PARC, and $s=0.5$ for generating long horizon motions on complex terrain using the final motion generator.}

\begin{tabular}{|c | c c c c|}
 \hline
 \textbf{Blending Coefficient} & \textbf{FWD} $\downarrow$ & \textbf{TPL} $\downarrow$ & \textbf{TCL} $\downarrow$ & \textbf{\%HJF} $\downarrow$ \\ [0.35ex] 
 \hline\hline
 0 & 0.908 & 40796 & 185.3 & 1.479 \\ 
 \hline
 0.25 & 0.776 & 7411 & 44.93 & 1.113 \\
 \hline
 0.5 & 0.571 & 4872 & 32.58 & 1.017 \\
 \hline
 0.65 & 0.596 & 179.6 & 9.763 & 2.730 \\
 \hline
 0.75 & 0.574 & 132.2 & 7.718 & 8.434 \\
 \hline
 1 & 0.537 & 129.8 & 6.751 & 54.82 \\
 \hline
\end{tabular}
\label{tbl:cfg-metrics}
\end{table}


\section{Discussion and Future Work}
In this work, we introduced PARC, a data-augmentation framework for training versatile physics-based terrain traversal controllers starting with only a small motion dataset. PARC enhances training by jointly optimizing a motion generation model and a physics-based motion tracking controller, with the two models generating data for each other in a synergistic process.
Once trained, the system allows simulated characters to navigate complex environments with agility, using enhanced motion generation and tracking from this iterative co-training approach.

PARC leverages the motion generation model and motion tracking controller to progressively expand the capabilities in the dataset, while mitigating physical artifacts by leveraging a physics simulation.
However, PARC does not fully eliminate the risk of the model developing unnatural behaviors. 
Exploring more sophisticated techniques to identify and filter out unnatural movements can potentially improve the realism of the synthesized motions.
Furthermore, our models are trained using procedurally generated terrain that lack the diversity and complexity of real world environments.
Enhancing our framework to accommodate more complex and realistic scenes will enable the models to synthesize behaviors better suited for interactions with more life-like environments.
Finally, our motion generator is not fast enough for real-time closed-loop planning, which is a necessary condition for video games and robotics.

\begin{acks}
We would like to thank Beyond Capture Studios for providing the parkour motion data. Funding for this work was provided by the National Research Council Canada, and NSERC via a Discovery Grant (RGPIN-2015-04843). We would also like to thank the anonymous reviewers for their constructive comments that enhanced the final version of this paper.
\end{acks}

\clearpage
\bibliographystyle{ACM-Reference-Format}
\bibliography{bib}

\clearpage
\appendix
\section{Terrain}

\label{app:terrain}
Our goal is to design a physics-based character controller capable of traversing diverse and complex terrain. This is however a very challenging task, and providing sufficient observations of the terrain to our character controller can be very memory intensive. Therefore, we apply some constraints to the terrain in PARC to simplify the task while still being able to achieve impressive results.

\subsection{Terrain Representation}

\begin{figure}
    \centering
    \includegraphics[width=1\linewidth]{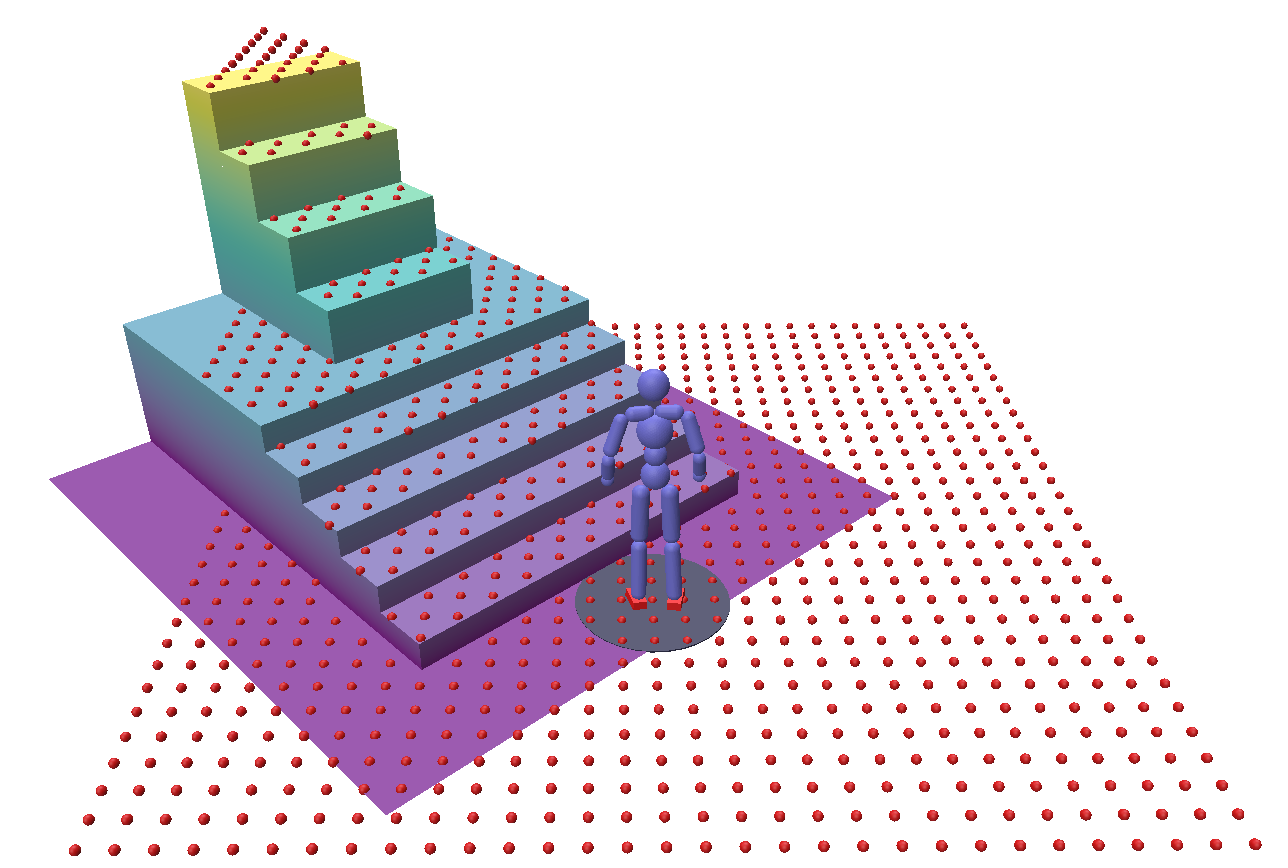}
    \includegraphics[width=0.45\linewidth]{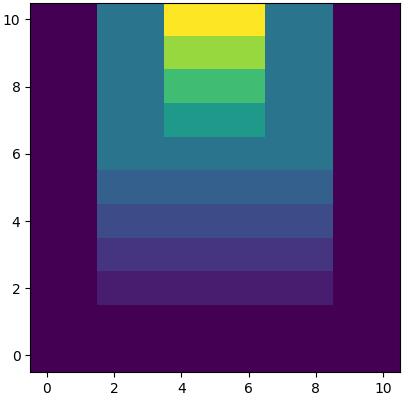}
    \includegraphics[width=0.45\linewidth]{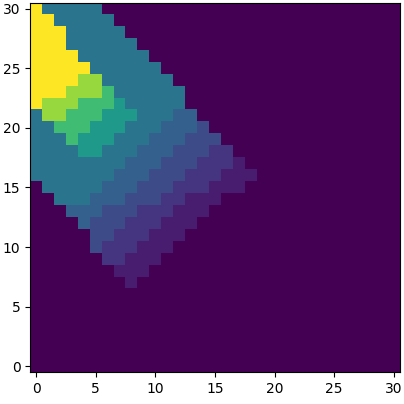}
    \caption{(Top) An example of a terrain used in our experiments, as well as a character standing on the terrain. The character's local heightmap is visualized as red points. (Bottom left) The global heightmap of the terrain. (Bottom right) The local heightmap $\rvh$ that is input to the motion generator.}
    \label{fig:terrain-example}
\end{figure}

Our terrains are represented as 2.5D grids.
Each grid index \( i, j \) corresponds to a box centered at a terrain point \( (x_0 + i \Delta x, y_0 + j \Delta y) \), where \( (x_0, y_0) \) represent the minimum box center coordinates, and \( \Delta x, \Delta y \) denote the box dimensions. In particular, we use $\Delta x = \Delta y = 0.4$m. The top surface of the box is located at \( \rvh(i, j) \) meters, while the bottom surface effectively extends to \( -\infty \) since our terrains are 2.5D. An example of a terrain is shown in Figure \ref{fig:terrain-example}.

\subsection{Terrain Generation}

\subsubsection{Random Boxes}

\begin{figure*}[htbp]
    \centering
    \includegraphics[width=0.24\linewidth]{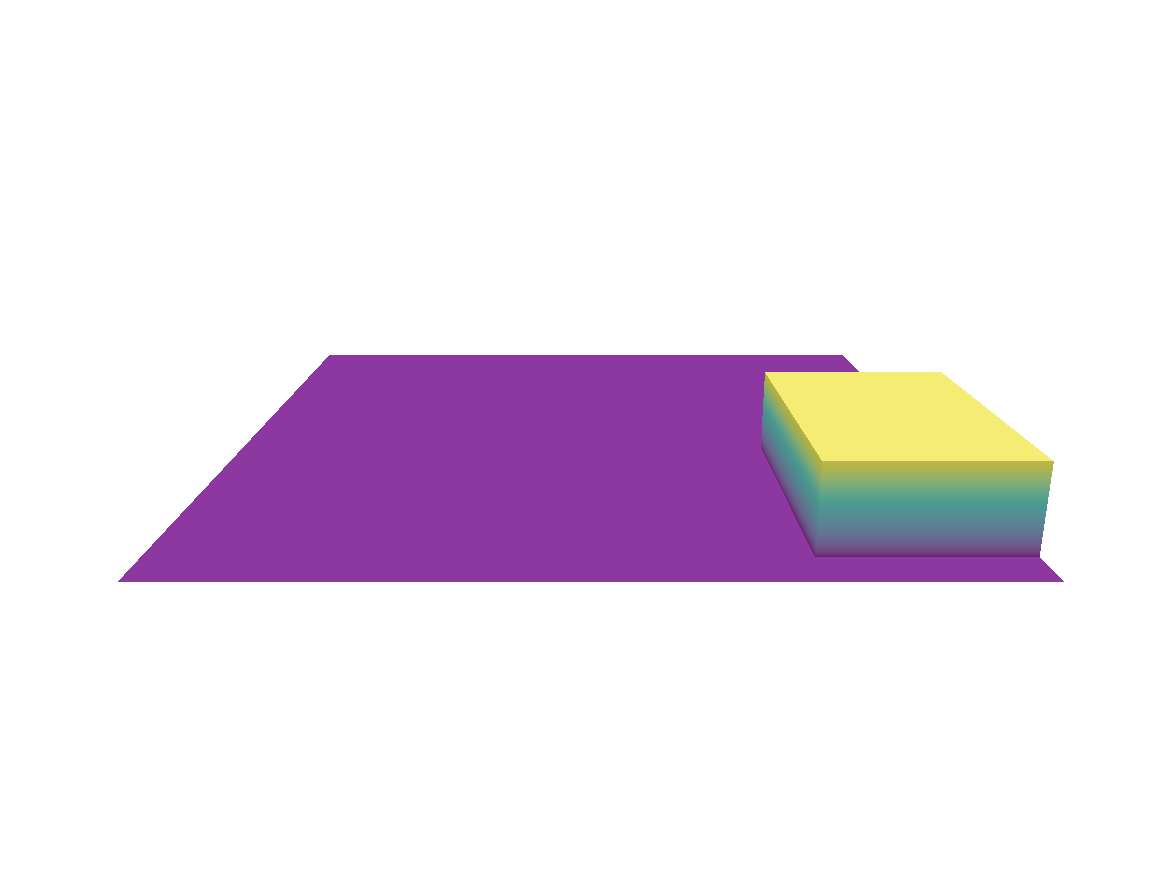}
    \includegraphics[width=0.24\linewidth]{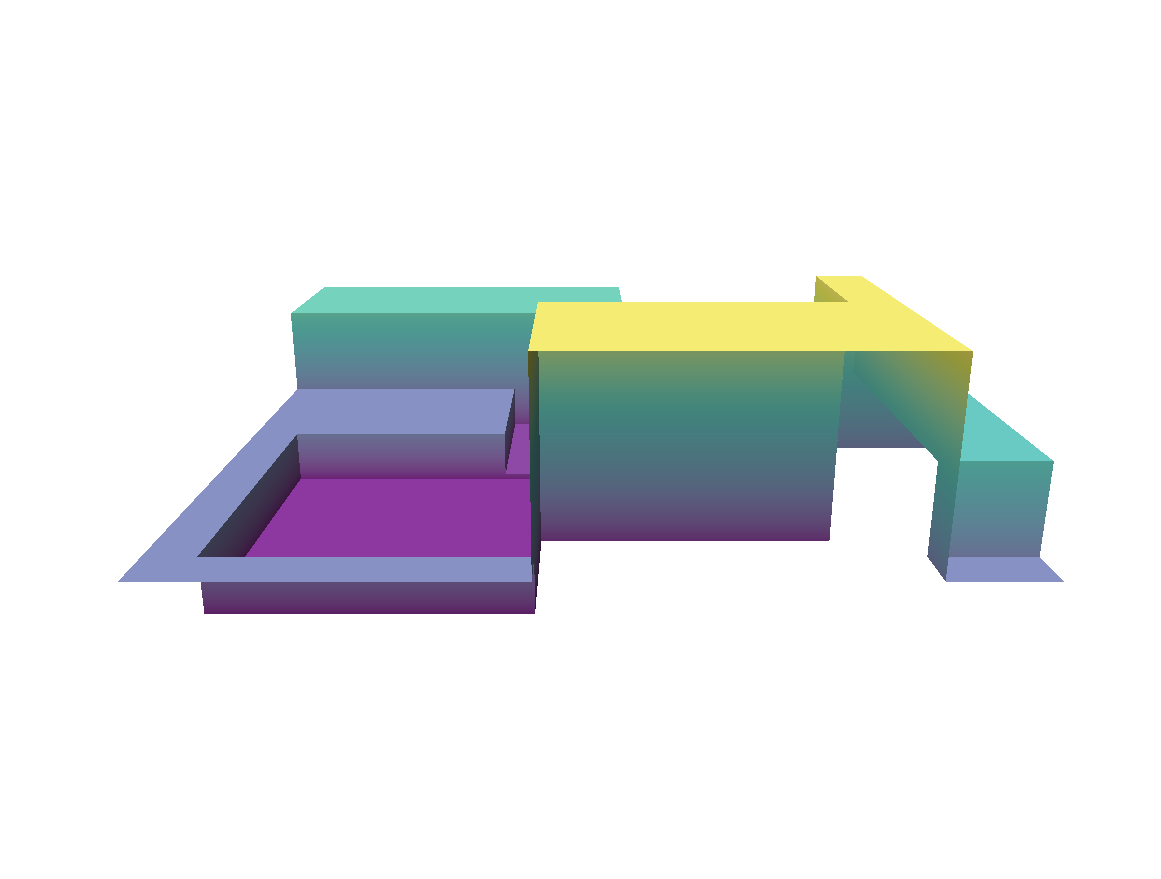}
    \includegraphics[width=0.24\linewidth]{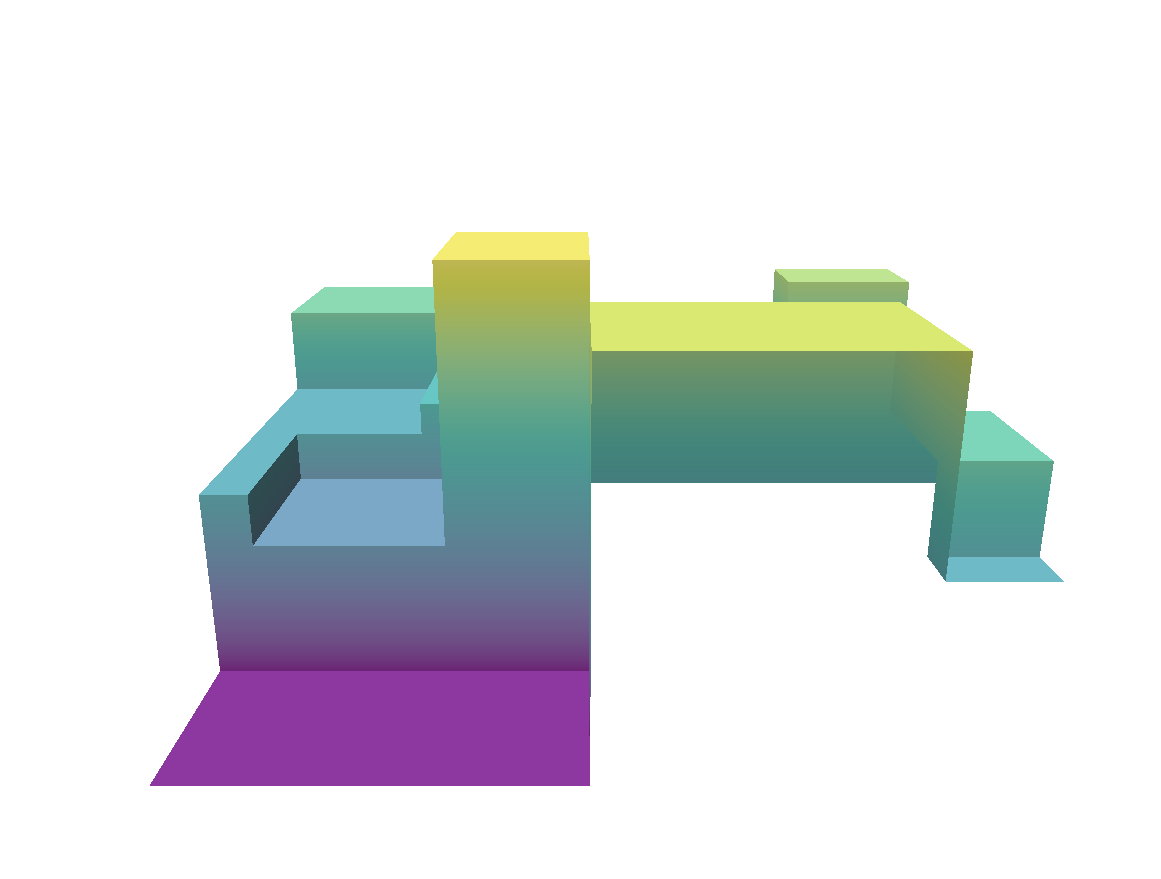}
    \includegraphics[width=0.24\linewidth]{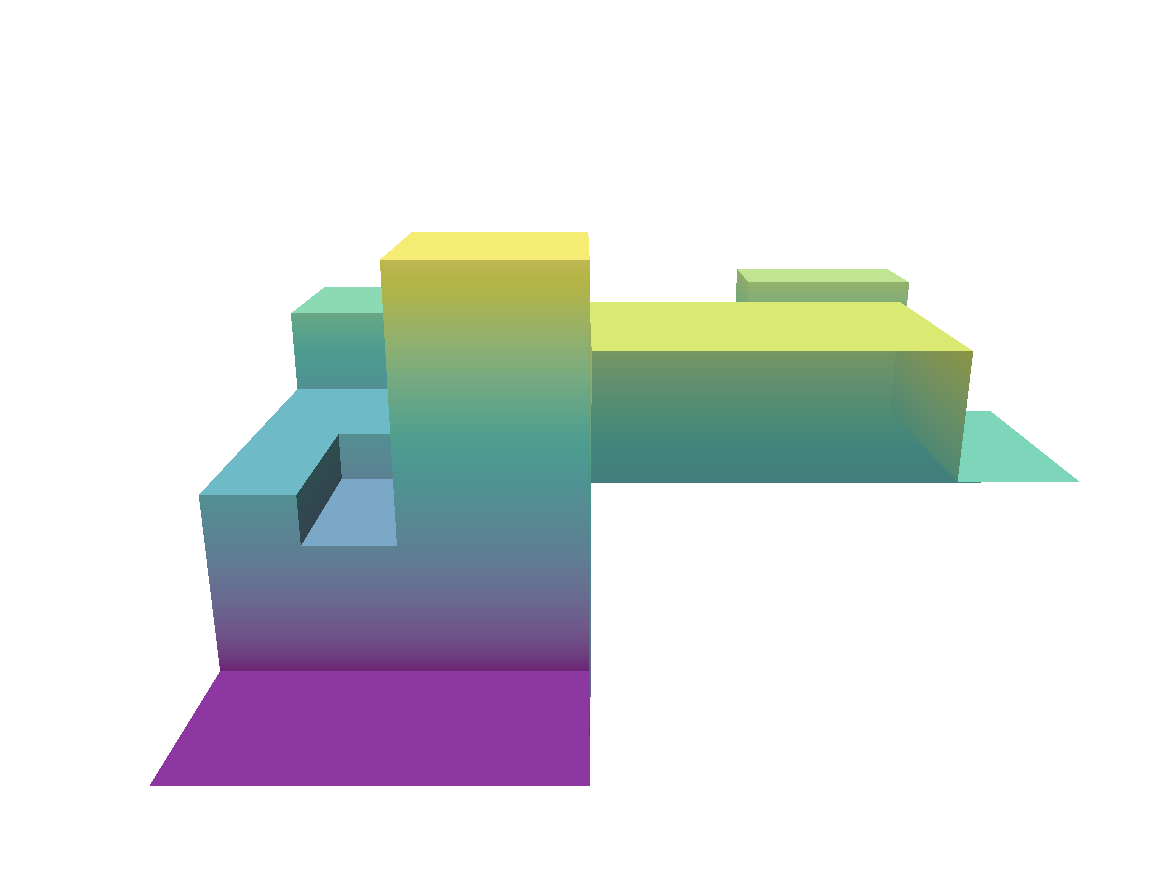}
    \includegraphics[width=0.24\linewidth]{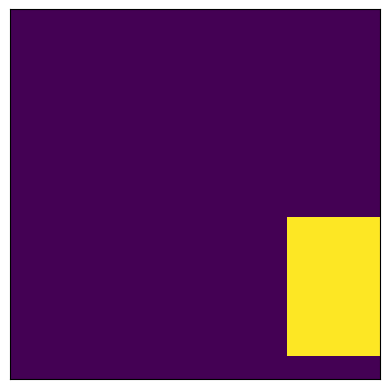}
    \includegraphics[width=0.24\linewidth]{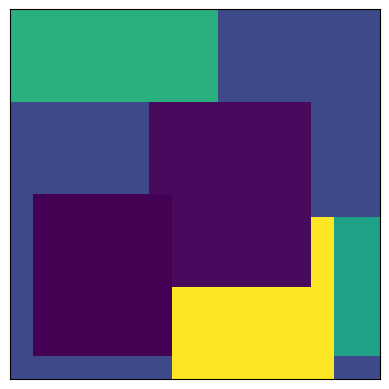}
    \includegraphics[width=0.24\linewidth]{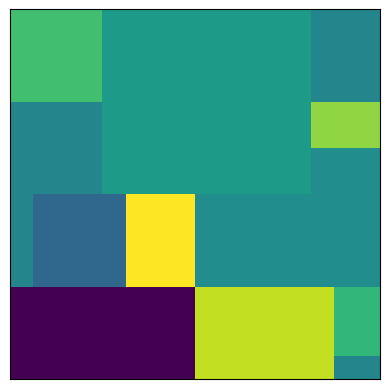}
    \includegraphics[width=0.24\linewidth]{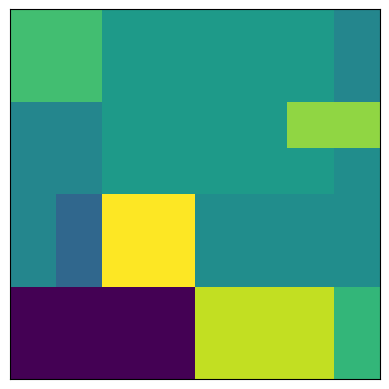}
    \caption{An illustration of our Random Boxes terrain generation method. From left to right: we begin with a single randomly generated box on a flat terrain. Next, we show terrains with five and ten randomly generated boxes, respectively, where each new box is added cumulatively. In the final step, we apply a sliding 2×2 max-pooling operation to the terrain heightmap, which removes narrow corridors and small platforms—producing smoother, more navigable terrain for the motion generator.}
    \label{fig:random-boxes-demo}
\end{figure*}

Given a terrain grid of size NxM, the Random Boxes method sequentially generates B boxes with random centers, widths, lengths, and heights. 
For generating new terrains in iteration 1 and 2 of PARC, we apply the Random Boxes method to a flat 16x16 grid, with box widths and lengths within 5-10 grid cells, and box heights within -2m to 2m. 
To simplify the terrain and make it easier to generate motions on, we ensure there are no thin 1 block gaps or walls. 
We do this by using a sliding 2x2 window with a stride of 2 to flatten grid cells within the window. 
The flattening is done by setting grid cells within a window to the maximum of height within the window.
An example of the Random Boxes method for generating a new terrain can be seen in Figure \ref{fig:random-boxes-demo}.


\subsubsection{Random Walk Terrain Generation Algorithm}
The random walk terrain generation algorithm is used to generate our test terrains. We do this to better measure the performance of our motion generator and motion tracker on terrains generated by algorithms different from the generation algorithms used for training. This algorithm generates paths with random walks on a flat input terrain. The height of a path is randomly determined. For our 100 randomly generated test terrains, we initialized 32x32 flat terrains, then generated 10 random walk paths with random heights.

\subsubsection{Random Terrain Slices}
\begin{figure}
    \centering
    \includegraphics[width=1.0\linewidth]{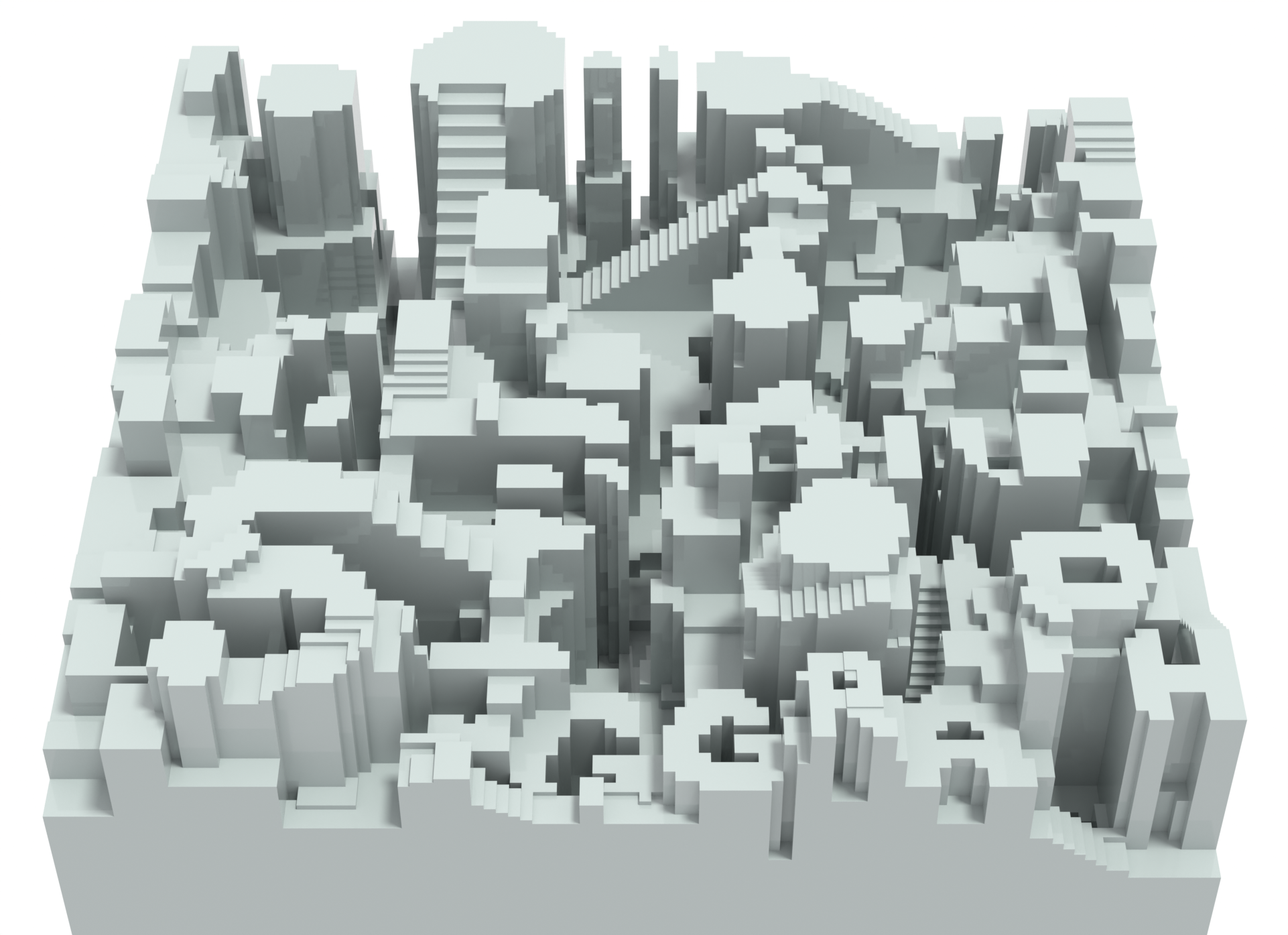}
    \caption{A 100x100 manually designed terrain used to help generate motions for the third iteration of PARC.}
    \label{fig:teaser-terrain}
\end{figure}

Given a large terrain, we can randomly select small slices of the terrain to run our path planner and motion generator on. We use a 100x100 manually designed terrain, shown in Figure \ref{fig:teaser-terrain}, and select random 16x16 slices to generate new motions on.

\subsection{Terrain Augmentation}
\begin{figure}
    \centering
    \includegraphics[width=1\linewidth]{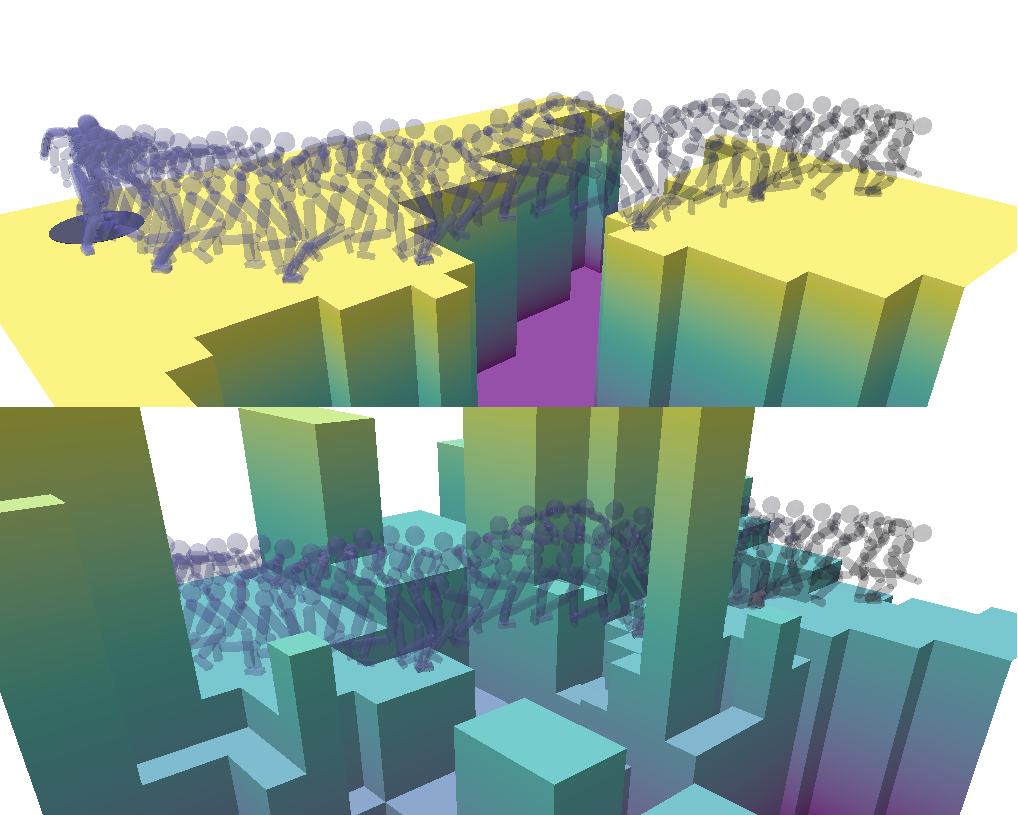}
    \caption{(Top) A running and jumping motion from the dataset as well as its associated terrain. (Bottom) The terrain augmented motion, where the terrain does not interfere with the original motion trajectory.}
    \label{fig:terrain-aug}
\end{figure}

Given a motion sequence and its associated terrain, we can alter the terrain without interfering with the motion, augmenting our dataset with more possible terrain observations. We use a simple a method by placing randomly rotated boxes on the terrain. In order for these boxes to make physical sense with the motion, we use a non-interfering augmented heightmap condition. Given the original motion sequence, it is possible to calculate upper and lower height bounds for each grid cell of the terrain heightmap such that the terrain does not intersect with any frame of the motion sequence. Then, after augmenting the terrain with randomly placed boxes, we clamp the terrain heights to be within the precomputed bounds. When augmenting local terrain heightmaps for training the motion generator, we apply an approximate non-interfering augmentend heightmap condition by assuming the local heightmap corresponds to a terrain geometry. An example of what these non-interfering terrain augmentations look like on a motion clip can be seen in Figure \ref{fig:terrain-aug}.

\section{Path Planning}
We use a custom A* \cite{1968-astar} algorithm to search for suitable paths between start and end points on terrains.
In order to apply A* to our task, we need to define the weighted graph using the terrain grid.
We also need to define a cost function for traversing the edges of the graph.

\subsection{Navigation Graph}

The graph for A* search is constructed as follows: 
\begin{itemize}
    \item Each cell has a directed edge connecting it to it's adjacent cells that are within an allowable height difference. We set the maximum difference to be 2.1 meters, which is slightly higher than the maximum height differences in the original dataset.
    \item To enable jumping over gaps, for each node we search for nodes within a certain jump radius and within a min and max jump height for each cell. If a node that satisfies the jump conditions exist, we add a directed jump edge.
    \item We also make sure not to add jump edges that intersect with walls by computing line box intersections for each potential jump edge.
\end{itemize}

\begin{figure*}[htbp]
    \centering
    \includegraphics[width=0.33\linewidth]{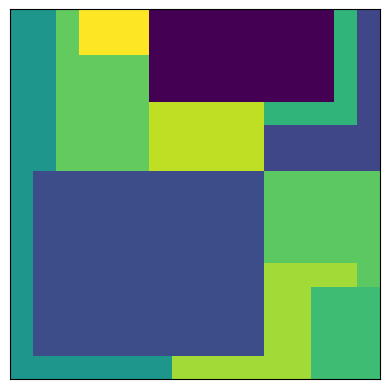}
    \includegraphics[width=0.33\linewidth]{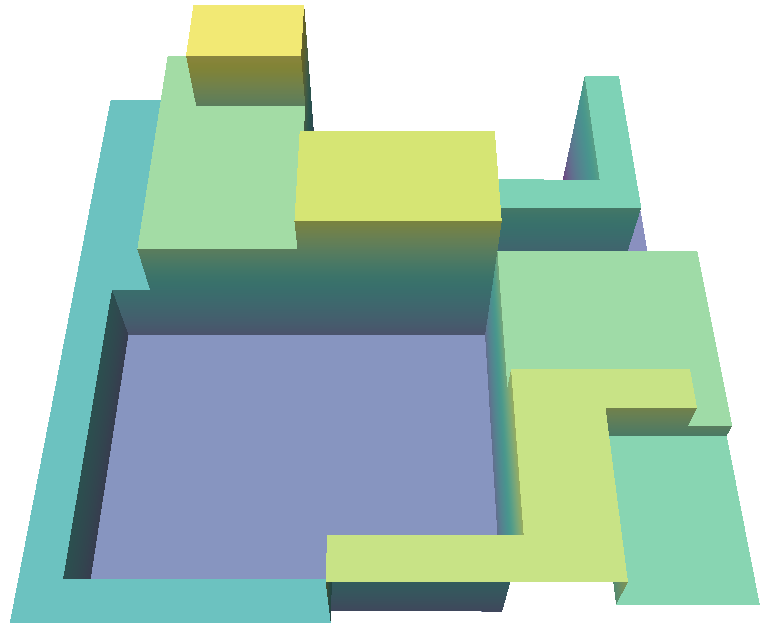}
    \includegraphics[width=0.33\linewidth]{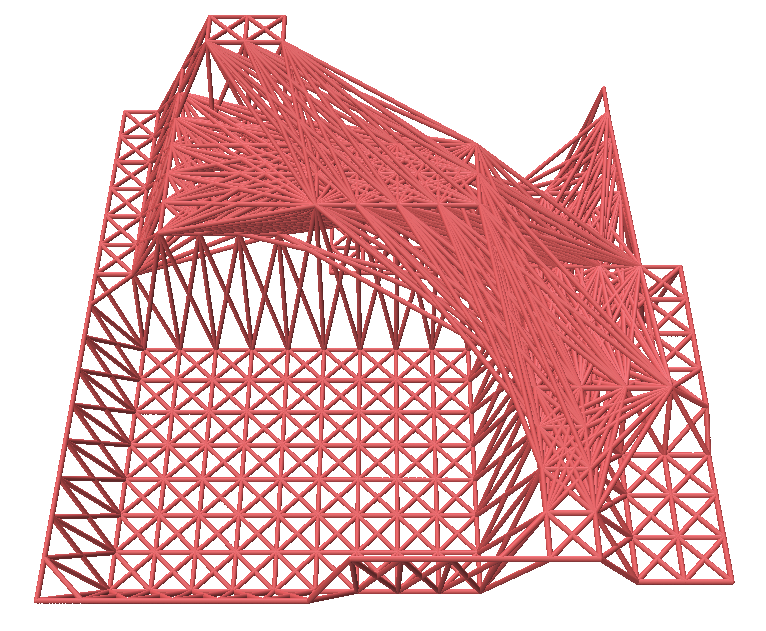}
    \caption{(Left) An example terrain heightmap grid. (Middle) The terrain's 3D visualization. (Right) The terrain's navigation graph, which is used for A* path planning.}
    \label{fig:path-graph}
\end{figure*}

\begin{figure*}[htbp]
    \centering
    \includegraphics[width=0.49\linewidth]{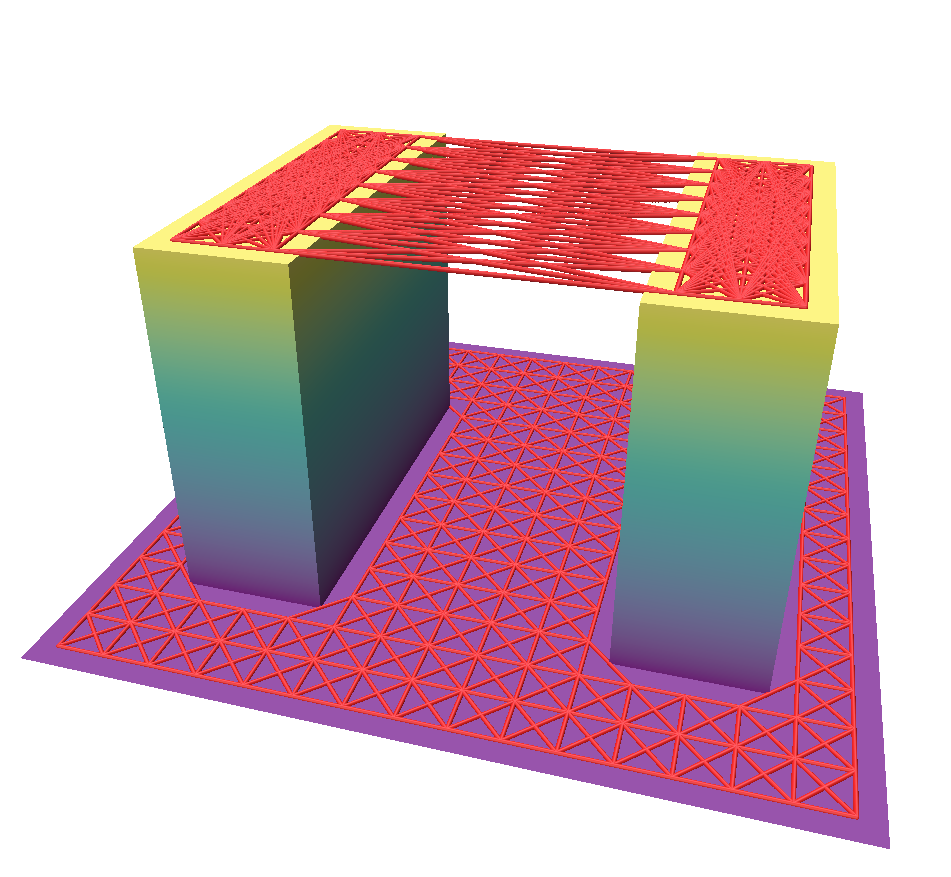}
    \includegraphics[width=0.49\linewidth]{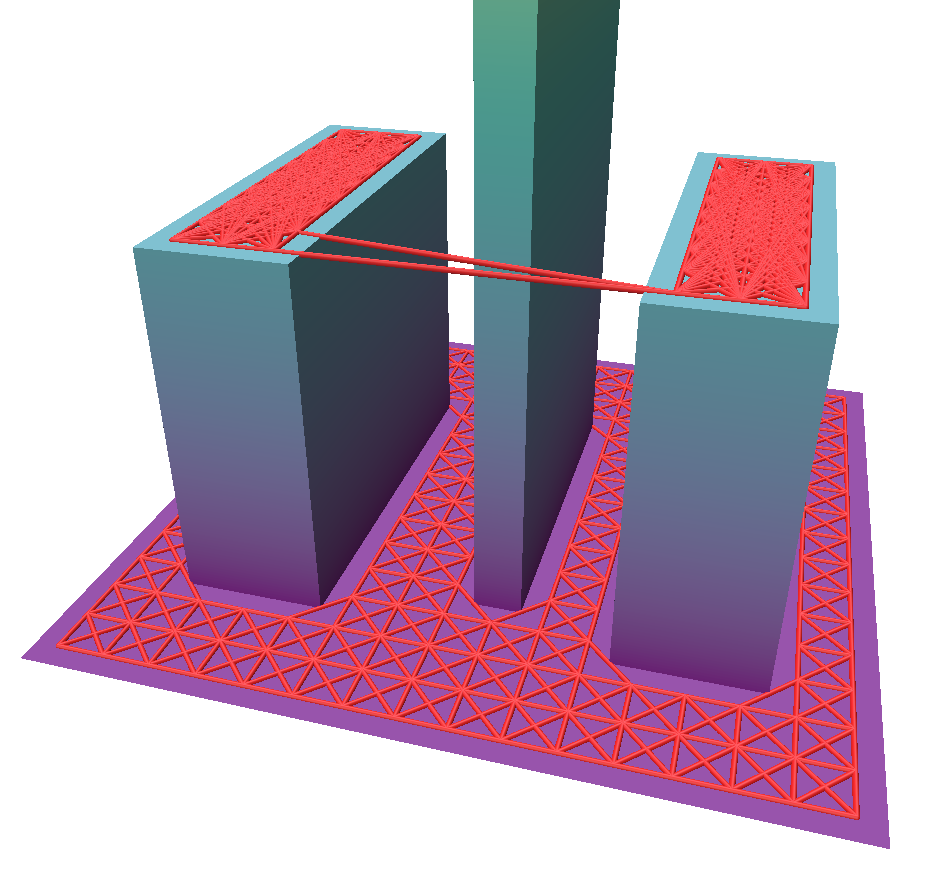}
    \caption{A visualization of the navigation graph with jump edges. (LEFT) We connect edges between cliff nodes within a jump radius, which are determined by having adjacent nodes that are at a much lower height. This creates connections in the graph that allows the path planner to jump across platforms. (RIGHT) Our navigation graph does not allow jump edges when a wall is interfering with the jump trajectory.}
    \label{fig:navgraph-jump}
\end{figure*}

A visualization of a terrain and it's navigation graph can be seen in Figure \ref{fig:path-graph}. A clear example of how jump edges interact with walls can be seen in Figure \ref{fig:navgraph-jump}.

\begin{figure*}[htbp]
    \centering
    \includegraphics[width=0.33\linewidth]{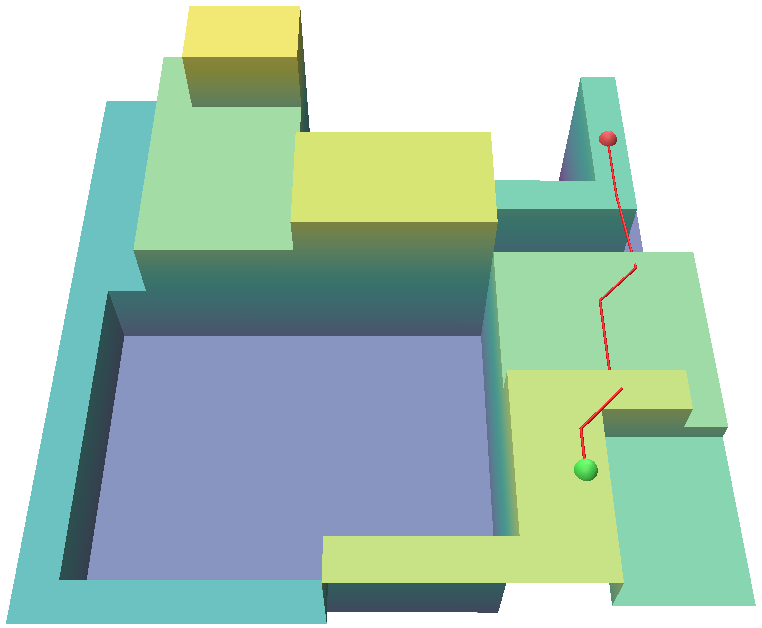}
    \includegraphics[width=0.33\linewidth]{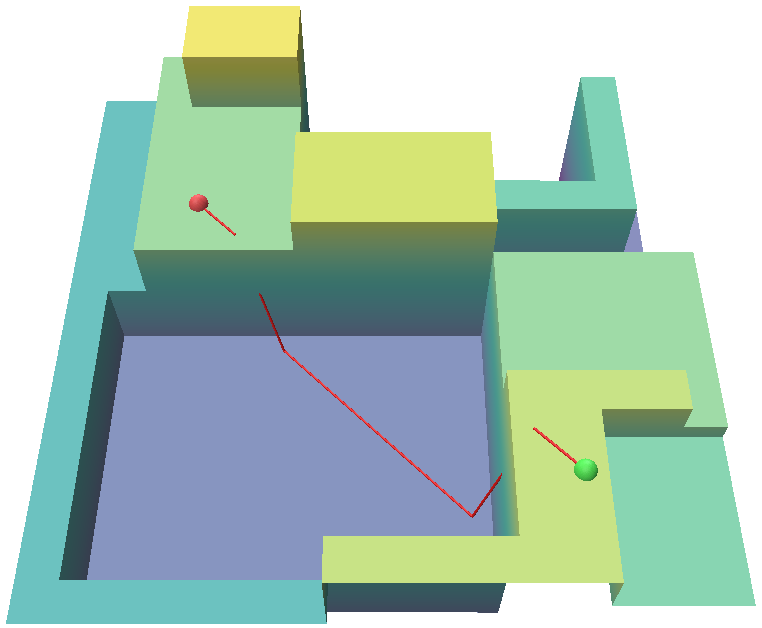}
    \includegraphics[width=0.33\linewidth]{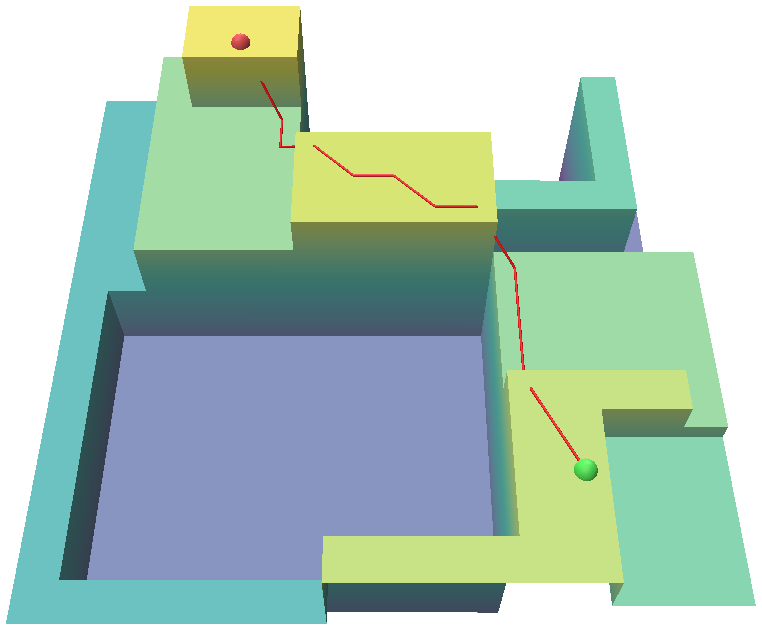}
    \caption{Examples of paths generated by our custom A* path planner.}
    \label{fig:astar-paths}
\end{figure*}

\subsection{Cost}
The cost to move along an edge is a function of the horizontal and vertical distance between the source node and potential target node. The horizontal and vertical distances can be weighted differently. In order to allow for more diversity in the generated paths, we also add a stochastic value to the cost function. In total, the cost function $g$ to move between the current node $\mathbf{x}_1 = (x_1, y_1, z_1)$ and a potential next node $\mathbf{x}_2 = (x_2, y_2, z_2)$ is:

\begin{equation}
    g(\mathbf{x_1}, \mathbf{x_2}) = w_{xy}((x_1-x_2)^2 + (y_1 - y_2)^2) + w_z(z_1 - z_2)^2 + X,
\end{equation}

where $X \sim U(c_{\text{min}}, c_{\text{max}})$ is a random variable to add stochasticity to the cost function. In our implementation, $w_{xy} = 1$, $w_{z} = 0.15$, $c_{\text{min}} = 0$, $c_{\text{max}} = 0.5$. The reason we use a higher horizontal distance weight is to prioritize short horizontal paths, therefore encouraging our path planner to make use of vertical terrain traversal skills such as climbing.

\subsection{Path Generation on New Terrains}
To generate a path for a potential new motion, we select a start and end node near the edges of randomly generated terrains and then run our custom A* algorithm. 
An example of a generated path on a terrain can be shown in Figure \ref{fig:astar-paths}.

\section{Heuristic Losses}
\label{app:losses}
We approximate geometric losses by computing signed distance functions between points sampled on the surface of the character and the terrain.

\subsection{Approximate Distance Fields}
Due to the way our terrain is constructed, it can be represented as a signed distance field. Each cell $(i, j)$ can be represented with a box centered at $(x, y) = (x_0 + i \Delta x, y_0 + j \Delta y)$ and with a top surface at $\mathbf{h}(i, j)$. We call the signed distance function to the terrain sdTerrain, which consists of the union of sdBox \cite{iqsdf} for each box in the terrain.

We approximate distance computations between the character body and terrain by using points sampled on the surface of the character, and then use sdTerrain to get the signed distances between the surface sampled points and the terrain.

\subsection{Terrain Penetration Loss}
Let $\rvp_i$ denote the $i^{th}$ point sampled on the body of the character. Then the terrain penetration loss is formally:
\begin{equation}
\label{eq:pen-loss}
    \gL_{\text{pen}} = \sum_{i=1}^{N_{\text{points}}} -\min(\text{sdTerrain} (\rvp_i), 0)
\end{equation}

\begin{figure}
    \centering
    \includegraphics[width=0.75\linewidth]{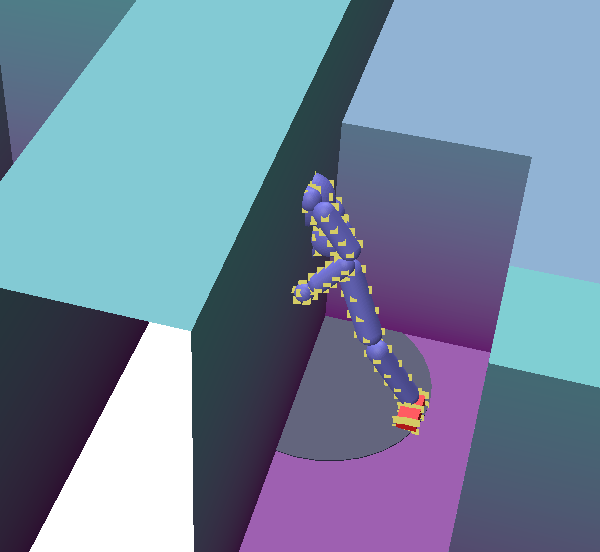}
    \includegraphics[width=0.75\linewidth]{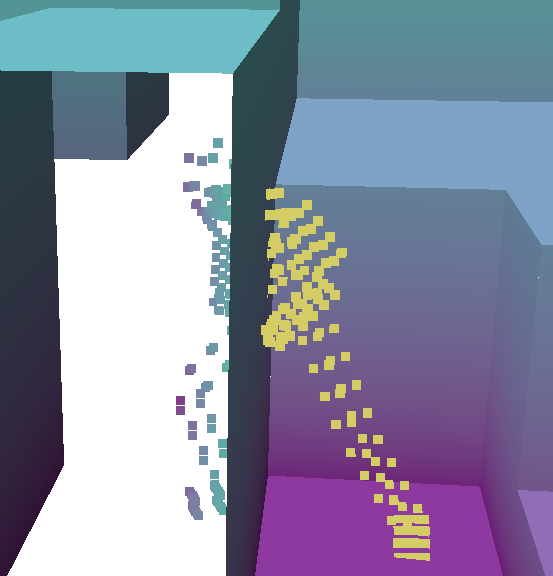}
    \caption{Visualizations of the approximate terrain penetration distances using points sampled on the surface of the character. The points are colored with a viridis color map, where darker represents more penetration.}
    \label{fig:terrain-pen}
\end{figure}

A visualization of the signed distances sdTerrain($\rvp_i$) on the points sampled on the character's body surface can be seen in Figure \ref{fig:terrain-pen}.

\subsection{Terrain Contact Loss}
We use the contact labels to determine when a body part is supposed to be in contact with the terrain. If a body is supposed to be in contact, then the distance between at least one of the sampled points on the body part and the terrain should be 0. Let $P(b)$ denote the set of points sampled on the surface of the $b^{th}$ body part. Let $\rvc_b$ denote the binary contact label for the $b^{th}$ body part. Then the contact loss for one motion frame is: 
\begin{equation}
\label{eq:contact-loss}
    \gL_{\text{contact}} = \sum_{b=1}^{N_{\textjoints}} \rvc_b \text{min}_{\rvp \in P(b)} |\text{sdTerrain}(\rvp)|
\end{equation}

\subsection{Jerk Loss}
We compute the third derivative of joint position, which is joint jerk, using finite differences. We then add a penalty when a frame has a jerk magnitude greater than a specified maximum value. Let $\rvp_b$ denote the position of the $b^{th}$ joint, and $\text{jerk}_{\text{max}}$ denote the max jerk we want to allow, then the jerk loss is:

\begin{equation}
\label{eq:jerk-loss}
    \gL_{\text{jerk}} = \sum_{b=1}^{N_{\textjoints}} \max(|\dddot{\rvp}_b| - \text{jerk}_{\text{max}}, 0)
\end{equation}






\section{Diffusion Model Details}
\begin{figure}
    \centering
    \includegraphics[width=\linewidth]{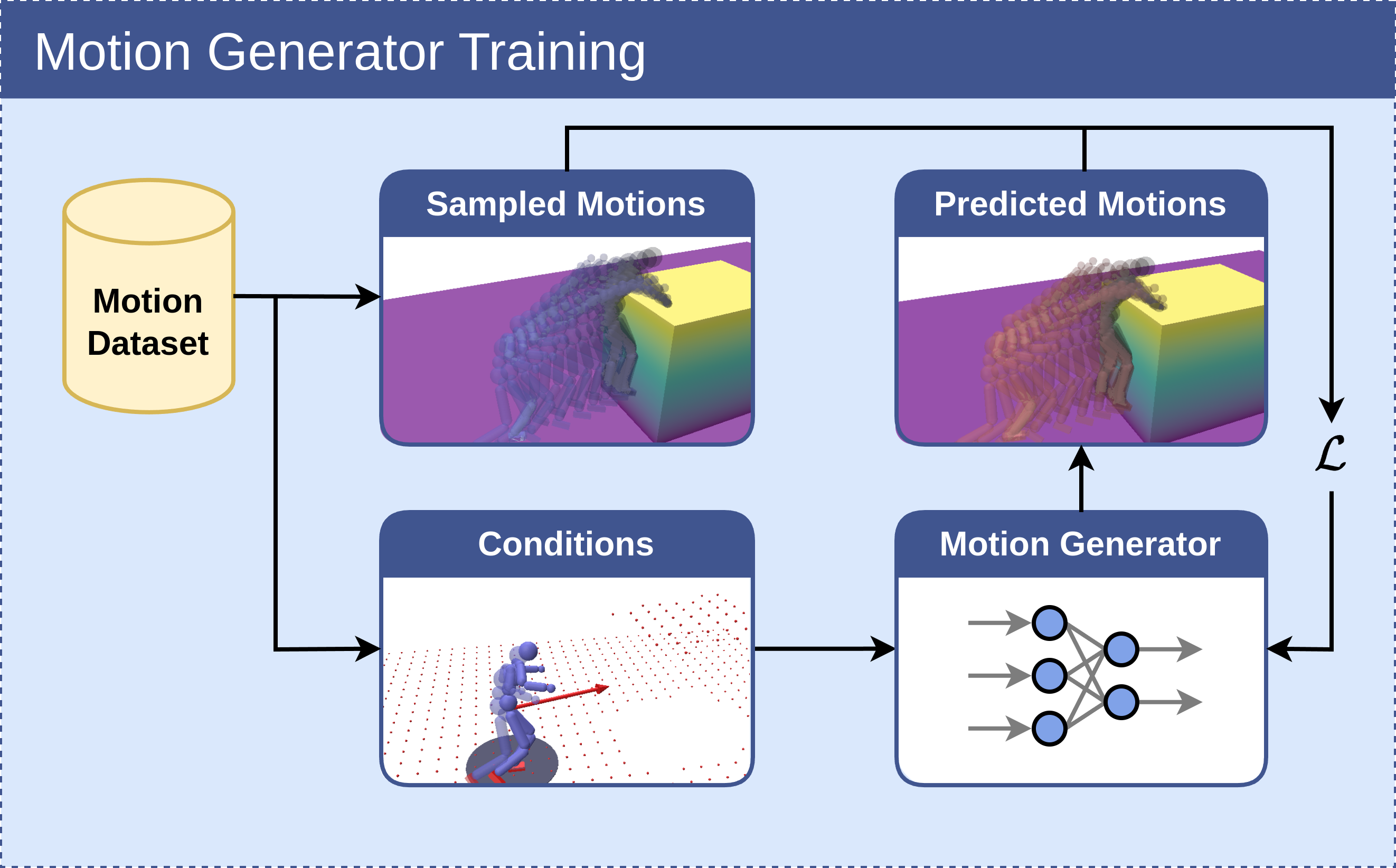}
    \caption{A diagram detailing a general recipe for training motion generators suitable for PARC.}
    \label{fig:motion-gen-training}
\end{figure}
In PARC, we use a diffusion model to represent the motion generator. However, we believe this design decision is not crucial for PARC, and we believe other models could work just as well. The general recipe for training kinematic motion generators for PARC is to sample motion sequences and their associated terrains from a dataset, and then use a reconstruction loss to predict the future frames given the past frames and terrain as contexts. Figure \ref{fig:motion-gen-training} shows this general recipe for training motion generators. In this section, we will go into more details about our particular diffusion model implementation for the motion generator.

\subsection{Diffusion Model Loss}
\label{app:mdm-losses}
The loss used for training the diffusion model is Eq \ref{eq:mdm-loss}. 
However, this loss would compute arithmetic distances between the rotations, which is not a valid metric for rotation differences. Instead, we split the reconstruction loss into positional $\rvp = \{ \rvp^{\textroot}, \rvp^{1:J}\}^{1:N}$ and rotational $\rvq = \{ \rvq^{\textroot}, \rvq^{1:J}\}^{1:N}$ components to compute the reconstruction loss appropriately, where $N$ is the number of motion frames and $J$ is the number of joints. We also extract the contact label component $\rvc = \{\rvc^{1:J}\}^{1:N}$. We denote $\hat{\rvp}_0$, $\hat{\rvq}_0$, and $\hat{\rvc}_0$ as the predicted positional component, rotational component, and contact labels extracted from $G(k, \rvx_k, \gC) = \hat{\rvx}_0$, respectively. Thus, the new reconstruction loss is:
\begin{flalign}
    &&\gL_{\text{rec}}(G) = \ddpmE \Bigl[||\rvp_{0} - \hat{\rvp}_0||^2\\
    &&+ ||\rvq_{0} \ominus \hat{\rvq}_0||^2 + ||\rvc_{0} - \hat{\rvc}_0||^2  \Bigr] \nonumber
\end{flalign}

We also include additional geometric losses. These losses are the velocity loss, joint position consistency loss, and terrain collision loss. The positional velocities $\dot{\rvp}$ are calculated using finite differences on the positional components of the motion frames. The angular velocities $\dot{\rvq}$ are calculated by first calculating quaternion finite differences, then converting them to exponential map form.
\begin{flalign}
    &&\gL_{\text{velocity}}(G) = \ddpmE \Bigl[ ||\dot{\rvp}_{0} - \hat{\dot{\rvp}}_0||^2 \\
    &&+ ||\dot{\rvq}_{0} - \hat{\dot{\rvq}}_0||^2 \Bigr]  \nonumber
\end{flalign}
The joint position consistency loss intends to connect the predicted 
joint positions and the joint positions computed using a forward kinematics function denoted as $FK(\cdot)$ on root position, root rotation, and joint rotations extracted from $\hat{\rvx}_0$.
\begin{equation}
    \gL_{\text{joint}}(G) = \ddpmE \left[||\hat{\rvp}_{0} - \text{FK}(\hat{\rvx}_0)||^2 \right]
\end{equation}
Finally, we use an approximate terrain penetration loss $\mathcal{L_\text{pen}}$, detailed in Eq \ref{eq:pen-loss}.

\subsection{DDIM Sampling}
When using a diffusion model that predicts the clean sample, we need to use a modified DDIM equation. Given a DDIM stride $d$ and an initial noise $x_K \sim \mathcal{N}(0, \mathbf{I})$, we can iteratively apply the following equation until we achieve the final predicted clean sample $\rvx_0$.
\begin{equation}
    \rvx_{k - d} = \Biggl(\sqrt{\bar{\alpha}_{k-d}} - \frac{\sqrt{\bar{\alpha}_k}\sqrt{1 - \bar{\alpha}_{k-d}}}{\sqrt{1 - \bar{\alpha}_{k}}} \Biggr) \hat{\rvx}_0 + \frac{\sqrt{1 - \bar{\alpha}_{k - d}}}{\sqrt{1 - \bar{\alpha}_{k}}}\rvx_k
\end{equation}

\section{Kinematic Motion Correction}
\label{app:kmc}

\begin{figure}
    \centering
    \includegraphics[width=1.0\linewidth]{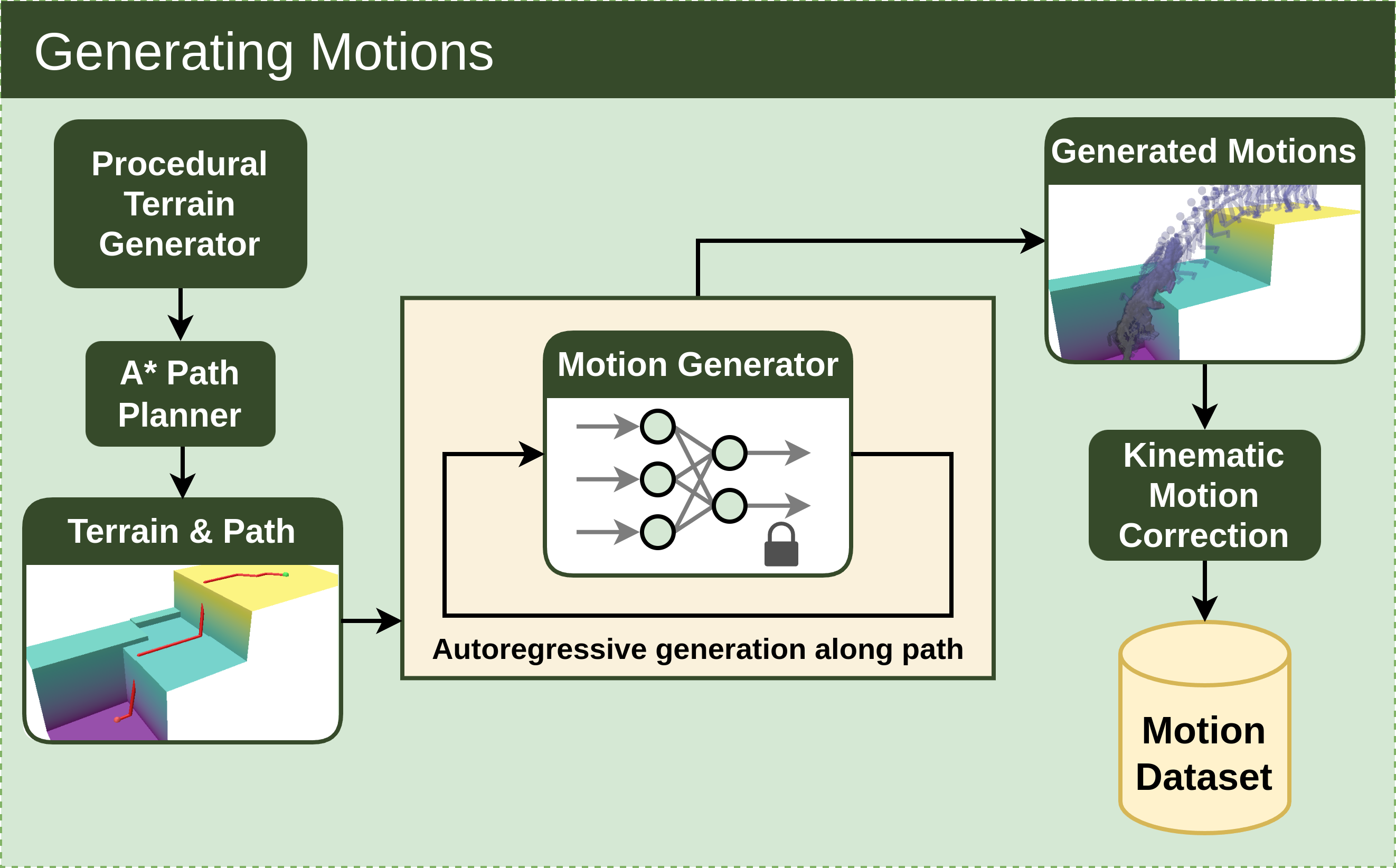}
    \caption{An overview of our kinematic motion generation pipeline. PARC's motion generation requires a terrain generation module, a path planning module, a trained motion generation model, and a kinematic motion correction module.}
    \label{fig:gen-motions}
\end{figure}

We use a collection of kinematic motion correction techniques to remove artifacts and enhance the quality of our generated motions, alleviating the difficulty of tracking using our motion tracking controller. These motion correction techniques are applied at the end of our kinematic motion generation pipeline, which Figure \ref{fig:gen-motions} provides an overview of.

\subsection{Selection Heuristic}
To help filter the outputs of our motion generator, we generate a batch of motions in parallel, then select the best motion based on a selection heuristic. The selection heuristic is combination of the terrain penetration loss in Equation \ref{eq:pen-loss}, terrain contact loss in Equation \ref{eq:contact-loss}, and path incompletion penalty. The path incompletion penalty is added to the loss if a generated motion is unable to reach the end of the path it is conditioned on after an amount of time decided by the user. The full selection heuristic that we used is:

\begin{equation}
\label{eq:selection}
    \gL_{\text{motion}} = \gL_{\text{pen}} + \gL_{\text{contact}} + 1000 (\text{reached end of path})
\end{equation}

\subsection{Kinematic Motion Optimization}
Motions produced by the motion generator can have various artifacts that make it particularly challenging to track in simulation. To address these, we optimize the motion sequence using a heuristic loss function designed to remove unnatural motion artifacts. The loss function consists of:
\begin{itemize}
    \item A regularization loss $\gL_{\text{reg}}$ to prevent the optimized motion sequence from changing its root position, root rotation, and joint rotations too much from the original motion sequence. This simply computes the difference between the current optimized variables and the original source variables.
    \item A terrain penetration loss that penalizes body-terrain penetration, as described in Equation \ref{eq:pen-loss}.
    \item A terrain contact loss that penalizes missed body-terrain contact based on each motion frame's contact labels, as described in Equation \ref{eq:contact-loss}.
    \item A jerk loss that penalizes body joint jerk greater than $1000 m/s^3$, as described in Equation \ref{eq:jerk-loss}.
\end{itemize}
The full loss that we optimize for generated motions before the motion tracking stage is:
\begin{equation}
    \gL = w_{\text{reg}}\gL_{\text{reg}} + w_{\text{pen}}\gL_{\text{pen}} + w_{\text{contact}}\gL_{\text{contact}} + w_{\text{jerk}}\gL_{\text{jerk}}.
\end{equation}

The weights we use are $w_{\text{reg}} = 1$, $w_{\text{pen}} = 1000$, $w_{\text{contact}} = 1000$, and $w_{\text{jerk}} = 1000$. We optimize using Adam \cite{kingma2014adam} with a step size of $0.001$ and 3000 iterations. The full implementation details are available at \url{https://github.com/mshoe/PARC}.

\section{Motion Tracking Details}
\label{rl-rewards}

\begin{figure}
    \centering
    \includegraphics[width=\linewidth]{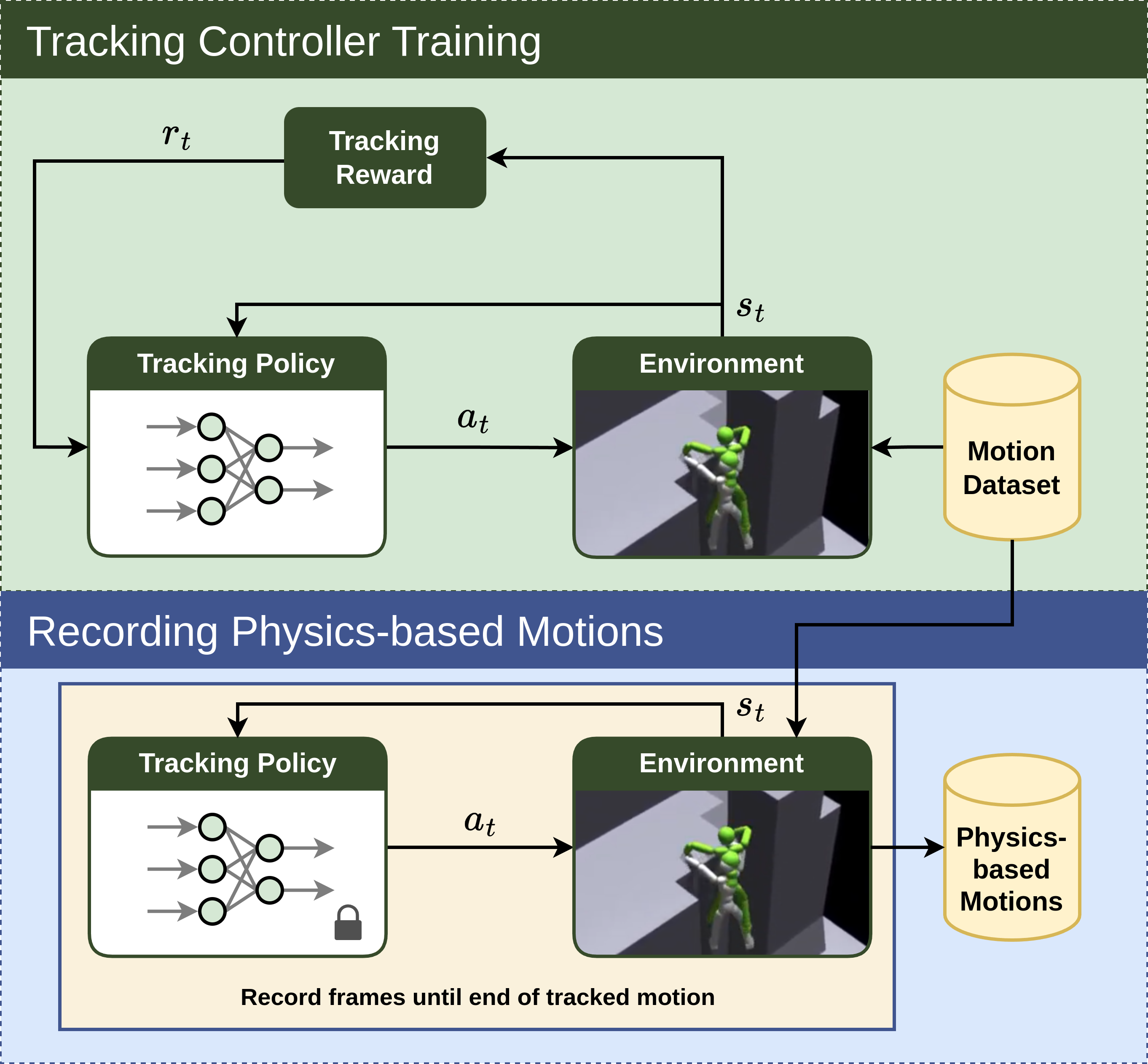}
    \caption{An overview of the reinforcement learning training method for the physics-based motion tracking controller, as well as the method for recording physics-based motions using the trained motion tracker.}
    \label{fig:tracker}
\end{figure}

An overview of training the motion tracker and recording physics-based motions can be seen in Figure \ref{fig:tracker}. The details of the reinforcement learning rewards and other motion tracking training techniques will be described in this section.

\subsection{Rewards}
We use a distance of $0.7$m for the pose termination criteria of each joint, except for the foot joints which we do not use a pose termination criteria on. This is to give the motion tracking agent more freedom in finding physically plausible solutions for tracking the kinematically generated motions.

The root position, root velocity, joint rotation, joint velocity, and key body position rewards are similar to \cite{peng2018deepmimic}. Given the character and the reference motion's joint rotations at time $t$, $\rvq_t ^j$ and $\hat{\rvq}_t ^j$, the joint rotation or "pose" reward is:
\begin{equation}
     r_t^{\text{pose}} = \exp{ \left[-0.25 \sum_j w_j ||\hat{\rvq}_t ^j \ominus \rvq_t ^j||^2 \right]}
\end{equation}
where $w_j$ is a tunable weight for the j$^{th}$ joint. The exact weights for character can be found in our publicly available code at \url{https://github.com/mshoe/PARC}.
Given the character and reference motion's local joint velocities,  $\dot{\rvq}_t ^j$ and $\hat{\dot{\rvq}}_t ^j$, the joint velocity reward is:
\begin{equation}
     r_t^{\text{pose velocity}} = \exp{ \left[ -0.01 \sum_j w_j ||\hat{\dot{\rvq}}_t ^j - \dot{\rvq}_t ^j||^2 \right] }
\end{equation}
Given the character and reference motion's root position and root rotations, $\rvp_t^{\textroot}$, $\rvq_t^{\textroot}$ and $\hat{\rvp}_t^{\textroot}$, $\hat{\rvq}_t^{\textroot}$, the root position reward is:
\begin{equation}
     r_t^{\text{root}} = \exp{ \left[-5 \left( ||\hat{\rvp}_t ^{\textroot} - \rvp_t ^{\textroot}||^2 + 0.1||\hat{\rvq}_t ^{\textroot} \ominus \rvq_t ^{\textroot}||^2 \right) \right]}
\end{equation}
Given the character and reference motion's root velocity and root angular velocity, $\dot{\rvp}_t^{\textroot}$, $\dot{\rvq}_t^{\textroot}$ and $\hat{\dot{\rvp}}_t^{\textroot}$, $\hat{\dot{\rvq}}_t^{\textroot}$, the root velocity reward is:
\begin{equation}
     r_t^{\text{root velocity}} = \exp{ \left[- \left( ||\hat{\dot{\rvp}}_t ^{\textroot} - \dot{\rvp}_t ^{\textroot}||^2 + 0.1||\hat{\dot{\rvq}}_t ^{\textroot} - \dot{\rvq}_t ^{\textroot}||^2 \right) \right]}
\end{equation}
Given the character and reference motion's key body positions, $\rvp_t^i$ and $\hat{\rvp}_t^i$, where $i$ denotes the key body index, the key body reward is:
\begin{equation}
     r_t^{\text{key}} = \exp{ \left[-10 \sum_i ||\hat{\rvp}_t ^i - \rvp_t ^i||^2 \right]}
\end{equation}
The key bodies in our experiments are the hands and feet of the humanoid character.

The contact label reward is both a reward and a penalty. It penalizes the agent when it's contact labels do not match the reference motion, but also rewards the agent when it does. This is to prevent the agent from learning motions that use unnatural contacts. Given reference contact labels $\hat{\rvc}_t$ and simulator computed contact labels for the character $\rvc_t$, the contact reward is:
\begin{equation}
    r_t^{\text{contact}} = \frac{1}{N_{\textjoints}} \sum_j \left[ \hat{\rvc}_t^j \rvc_t^j  - (1 - \hat{\rvc}_t^j) \rvc_t^j \right]
\end{equation}

The full tracking reward which is a weighted sum of the previously described rewards is:
\begin{flalign}
    &&r_t = 0.5 r_t^{\text{pose}} + 0.1 r_t^{\text{pose velocity}} + 0.15 r_t^{\textroot} + 0.1 r_t^{\text{root velocity}}\\
    && + 0.15 r_t^{\text{key}} + r_t ^{\text{contact}} \nonumber
\end{flalign}

\subsection{Prioritized State Initialization}
Training a tracking controller on a large motion dataset, particularly one comprising numerous contact-rich parkour motions, presents a significant multi-task learning challenge for reinforcement learning (RL) agents. Sampling motion clips uniformly across the dataset often results in an imbalance, where challenging motion clips receive disproportionately fewer samples. This issue is exacerbated by the use of early termination, as difficult motions that consistently fail are further deprived of sampling opportunities, while easier motions with lower failure rates consume a disproportionate share of resources.
In our framework, we track the failure rates of individual motion clips and incorporate these rates as sampling weights within a multinomial distribution. 
The sampling weights are used to sample a reference motion whenever an environment is reset. A motion clip is considered a failure if the agent fails to meet the pose termination criterion before completing the clip. By leveraging failure rates as sampling weights, PARC ensures that both challenging and unlearned motions receive proportionally more samples compared to easier or already mastered motions. To mitigate the risk of catastrophic forgetting, a minimum sampling weight of 0.01 is enforced for all motion clips, ensuring every motion continues to be sampled throughout training. Prior work uses variations of this technique known as prioritized state initialization \cite{won2019bodyshapevariation, 2019-predict-and-simulate, xie2021soccer, tessler2024maskedmimic}. 

\begin{figure}
    \centering
    \includegraphics[width=1.0\linewidth]{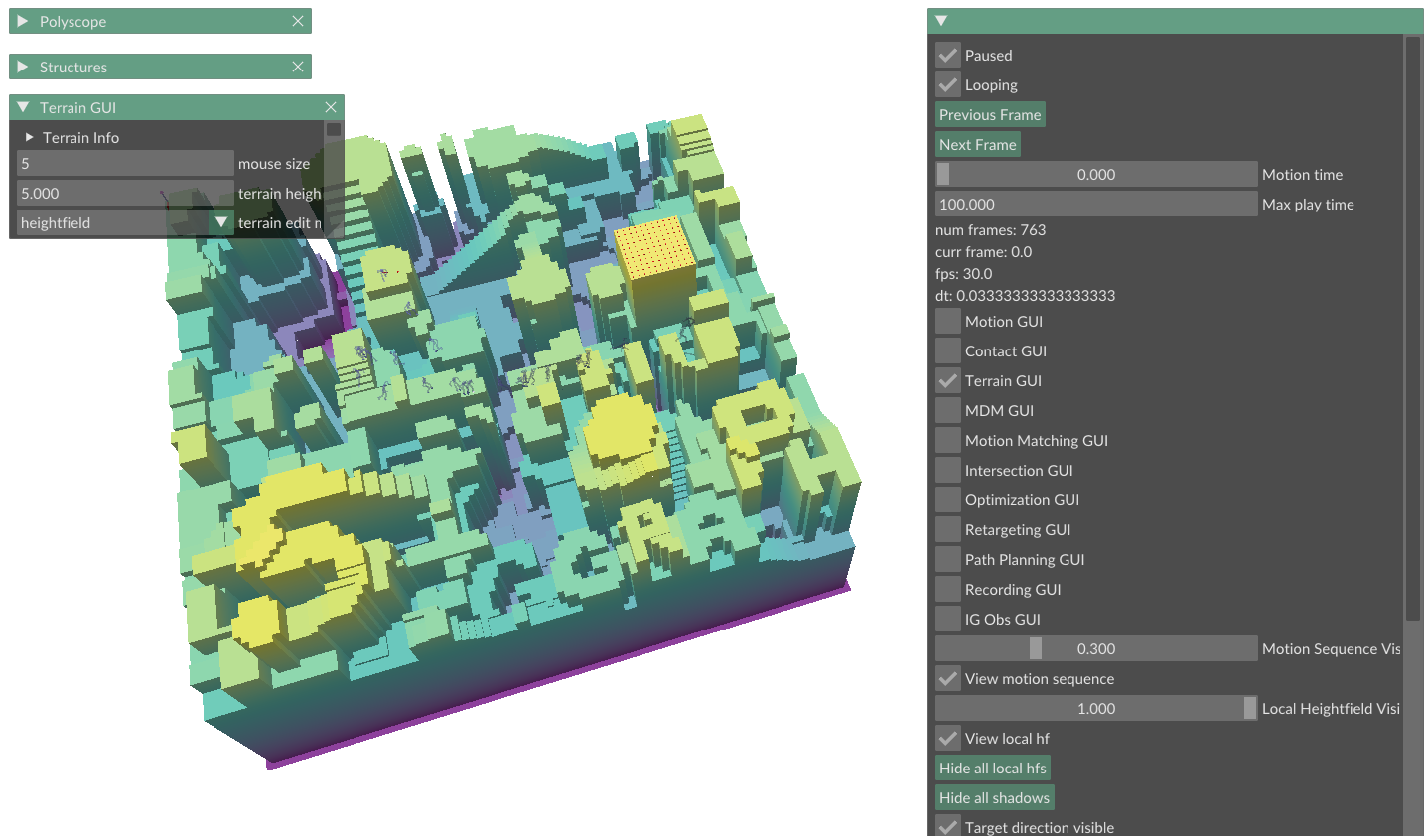}
    \caption{Our visualization, motion editing, terrain editing, and motion generation tool.}
    \label{fig:polyscope}
\end{figure}

\section{Visualization and Editing Tool}
In order to help both artists and researchers to use our work, we wrote a visualization tool using Polyscope \cite{polyscope} with motion and terrain editing functionalities. Our tool also allows users to draw their own paths or use our A* path planner, then use our motion generator to autoregressively generate a motion along the path. An image of our program can be seen in Figure \ref{fig:polyscope}.

\end{document}